\documentclass[%
reprint,
superscriptaddress,
amsmath,amssymb,
aps,
pra,
]{revtex4-2}

\usepackage{graphicx}
\usepackage{dcolumn}

\usepackage{dsfont}
\usepackage[capitalise,sort,noabbrev]{cleveref}
\usepackage{csquotes}
\usepackage{graphicx}
\usepackage{color}
\usepackage{asypictureB}

\begin{asyheader}
	usepackage("amsmath");
\end{asyheader}

\usepackage{tikz}

\usetikzlibrary{positioning,arrows,decorations.markings,decorations.pathreplacing,shapes.geometric,fit,calc,shadings}
\usepackage[miktex]{gnuplottex}

\usepackage{pgfplots}
\pgfplotsset{compat=1.7}

\pgfplotsset{
	legend image with text/.style={
		legend image code/.code={%
			\node[anchor=center] at (0.3cm,0cm) {#1};
		}
	},
}

\newcommand{\bra}[1]{\left\langle #1 \right|}
\newcommand{\ket}[1]{\left| #1\right\rangle}

\newcommand{\abs}[1]{\vert #1 \vert}
\newcommand{\abssq}[1]{\vert #1 \vert^2}
\newcommand{\idone}{\hat{\mathds{1}}}

\newcommand{\hvsigma}{\hat{\boldsymbol{\sigma}}}
\newcommand{\vsk}{\vec{s}_{K}}
\newcommand{\vtk}{\vec{t}_{K}}

\renewcommand{\vec}[1]{\boldsymbol{#1}}
\DeclareMathOperator\tr{tr}
\DeclareMathOperator\diag{diag}

\DeclareMathOperator{\arccot}{arccot}

\begin{document}

	\title{Geometric interpretation of the CHSH inequality of non-maximally entangled states}

	\author{Johannes Seiler}
	\affiliation{Institut f{\"u}r Quantenphysik \& Center for Integrated Quantum Science and Technology ($\mathrm{IQ^{ST}}$), Universit{\"a}t Ulm, D-89069 Ulm, Germany}
	\email{johannes.seiler@uni-ulm.de}
	\author{Thomas Strohm}%
	\affiliation{%
	Corporate Research, Robert Bosch GmbH, D-71272 Renningen, Germany	}%

	\author{Wolfgang P. Schleich}
	\affiliation{Institut f{\"u}r Quantenphysik \& Center for Integrated Quantum Science and Technology ($\mathrm{IQ^{ST}}$), Universit{\"a}t Ulm, D-89069 Ulm, Germany}%
	\affiliation{Institute of Quantum Technologies, German Aerospace Center (DLR), S{\"o}flinger Str.~100, D-89077 Ulm, Germany}%
	\affiliation{%
	Hagler Institute for Advanced Study, Institute for
	Quantum Science and Engineering (IQSE), and Texas A{\&}M AgriLife
	Research, Texas A{\&}M University, College Station, TX 77843-4242, USA	}%
		
	\date{\today}
	
	\begin{abstract}
	We show that for pure and mixed states the problem of maximizing the correlation measure in the CHSH inequality reduces to maximizing the perimeter of a parallelogram enclosed by an ellipse characterized by the entanglement contained in the bipartite system. Since our geometrical description is also valid for a non-maximally entangled state we can determine the corresponding optimal measurements.
	\end{abstract}
	
	\maketitle
	
	
\section{Introduction}
The Einstein-Podolsky-Rosen situation \cite{Einstein1935} represents a paradigm of quantum mechanics \cite{Bohr1935} and brings out the important role of entanglement \cite{Schrodinger1935}. This feature leads to a violation of the celebrated Bell inequality \cite{Bell1964,Brunner2014,scully1997quantum} or the Clauser-Horn-Shimony-Holt (CHSH) inequality \cite{Clauser1969}. In this article we provide a geometric interpretation of the quantum mechanical correlation measure central to the CHSH inequality.

Over the years Bell's inequality has been tested in many remarkable experiments \cite{Freedman1972,Fry1976,Aspect1982,Weihs1998,Hensen2015,Giustina2015,Shalm2015}, ever more strictly supporting quantum mechanics and its predictions. The key ingredient is hereby the entanglement of the underlying systems. Indeed, in order to violate the CHSH inequality the state has to be entangled \cite{Popescu1992}, and on the other hand, every entangled two-qubit state leads to a violation \cite{Gisin1991}, provided appropriate measurements are chosen. 

In fact, the optimal measurement depends crucially on the entanglement of the system. For example, the measurement settings that lead to the maximal violation of the CHSH-inequality for a given entangled state can fulfill the inequality for a state with different entanglement. In order to obtain a violation, one therefore has to adapt the measurement strategy to the state \cite{Popescu1992}.

In recent years, the connection between Bell inequalities and entanglement has been exploited in many  
quantum informational tasks, such as quantum key distribution \cite{Ekert1991} or quantum random number generators \cite{Pironio2010,Acin2012}.
Unfortunately, in real life experiments, the state is never maximally entangled, nor necessarily pure. Knowing which measurements lead to the maximal violation of the underlying Bell inequality for a given entanglement can therefore increase the efficiency of such tasks. 

In this article, we consider the CHSH inequality for non-maximally entangled states and introduce in Section \ref{sec:formulation} the state, the measurement operators and the expectation value of interest. In Section \ref{sec:geo}, we present a geometrical interpretation of the correlation function bounded by the CHSH inequality, where we connect this expectation value with the perimeter of a parallelogram caught by an ellipse whose semi-axes depend on the strength of the entanglement parametrized by the concurrence. This interpretation highlights the crucial role of the entanglement in violating the CHSH inequality. 
We then use this geometric argument in Section \ref{sec:optimization} to analytically determine all optimal measurement strategies for a given entangled pure two-qubit state.  
We finally briefly discuss the extension of our geometric interpretation to mixed states in Section \ref{sec:mixed}.
We conclude in Section \ref{sec:conclusionsandoutlook} by summarizing and providing an outlook. In order to keep our article self-contained but focused on the main ideas, we have included additional material in two appendices. In Appendix \ref{app:derivation} we present the analytical optimization of the measurement strategies, while in Appendix \ref{app:orthogonality} we prove that these strategies are found geometrically, when demanding that the tangents on the ellipse in the corners of the parallelogram are orthogonal.  

\section{Formulation of the problem}\label{sec:formulation}
In this section we briefly recall the essential ingredients of the CHSH inequality by first defining the corresponding quantum mechanical expectation value in terms of measurement operators. In order to lay the groundwork for our geometrical interpretation of the CHSH inequality, we then cast this expectation value in terms of the measurement directions and the correlation matrix. We conclude this section by presenting the Tsirelson bound.

Throughout this article, we consider the pure two-qubit state 
\begin{equation}\label{eq:Psi-Schmidt}
	\ket{\Psi} \equiv \sqrt{\frac{1+\sqrt{1-\mathcal{C}^{2}}}{2}} \ket{0}_{A}\ket{1}_{B} - \sqrt{\frac{1-\sqrt{1-\mathcal{C}^{2}}}{2}} \ket{1}_{A}\ket{0}_{B} ,
\end{equation}
in the Schmidt basis \cite{Nielsen2001}, where $ \ket{0} $ and $ \ket{1} $ are the eigenstates of the Pauli operator $ \hat{\sigma}_{z} $ with the eigenvalues $ 1 $ and $ -1 $, respectively. Here, the concurrence $ \mathcal{C} $ with $ 0 \leq \mathcal{C} \leq 1 $, is a measure of the entanglement between the two subsystems. For $ \mathcal{C}=0 $ the state is separable, while for $ \mathcal{C}=1 $ it is a maximally entangled Bell state.

On each subsystem we perform one of two measurements. In the subsystem $ A $, we either employ the operator \begin{equation}\label{eq:def-QandR}
	\hat{Q} \equiv \vec{q}\cdot \hvsigma_{A} \quad \mathrm{or} \quad \hat{R} \equiv \vec{r}\cdot \hvsigma_{A},
\end{equation}
with the three-dimensional unit vectors $ \vec{q}$ and $\vec{r} $ describing the measurement directions on the Bloch sphere, and $ \hvsigma_{A} $ denotes the vector formed by the Pauli matrices $ \hat{\sigma}_{x} $, $ \hat{\sigma}_{y} $ and $ \hat{\sigma}_{z} $. 

Analogously on the subsystem $ B $ we either make the measurement 
\begin{equation}\label{eq:def-SandT}
	\hat{S} \equiv \vec{s}\cdot \hvsigma_{B} \quad \mathrm{or} \quad \hat{T} \equiv \vec{t}\cdot \hvsigma_{B},
\end{equation} 
with the three-dimensional unit vectors $ \vec{s}$ and $\vec{t} $, and $ \hvsigma_{B} $ is again the vector of the Pauli matrices. The measurements on both subsystems are performed independently of each other. 

We are interested in the expectation value
\begin{equation}\label{eq:S-avg-Operators}
	\mathcal{S} \equiv \langle \hat{Q}\otimes \hat{S} \rangle- \langle \hat{Q}\otimes \hat{T} \rangle + \langle \hat{R}\otimes \hat{S} \rangle + \langle \hat{R}\otimes \hat{T} \rangle
\end{equation}
which appears in the CHSH inequality and is a measure for the correlation between the two subsystems \cite{Clauser1969}. 

When we substitute the state $ \ket{\Psi} $ defined by \cref{eq:Psi-Schmidt} into \cref{eq:S-avg-Operators}, and use the representations \cref{eq:def-QandR,eq:def-SandT} of the measurement operators we get the representation 
\begin{equation}\label{eq:S-Kqrst}
	\mathcal{S} = \vec{q} K\cdot(\vec{s} - \vec{t}) + \vec{r} K\cdot (\vec{s} +  \vec{t})
\end{equation}
in terms of multiplications between the Bloch vectors $ \vec{q},\vec{r},\vec{s},\vec{t} $ of the measurements and the $ 3\times 3 $ correlation matrix
\begin{equation}\label{eq:K-operators}
	K \equiv \bra{\Psi}\hvsigma_{A}\otimes \hvsigma_{B}\ket{\Psi}.
\end{equation}
This matrix depends only on the state $ \ket{\Psi} $ of the complete system. For the state $ \ket{\Psi} $ of \cref{eq:Psi-Schmidt} the correlation matrix is \cite{Seiler2020} the diagonal matrix
\begin{equation}\label{eq:K-diag}
	K = -\diag(\mathcal{C},\mathcal{C},1)
\end{equation}
and only depends on the concurrence $ \mathcal{C} $ and thus on the entanglement of the state. The diagonal form of the correlation matrix, \cref{eq:K-diag}, is not a coincidence, but the result of defining the state $ \ket{\Psi} $ in the Schmidt basis, in which the state is invariant under rotations around the $ z $-axis.  

For a mere classical system, that is one which obeys a local hidden variable theory, it is well known that the correlation measure is bounded by the CHSH inequality
\begin{equation}\label{eq:CHSH-inequality}
	\abs{\mathcal{S}} \leq 2.
\end{equation}
This bound can be violated by an entangled quantum mechanical state, and in fact, for any pure two-qubit quantum state the correlation measure $ \mathcal{S} $ is bounded \cite{Cirelson1980,Tsirelson1987} by
\begin{equation}\label{eq:Tsirelson-bound}
	\abs{\mathcal{S}}\leq 2\sqrt{2},
\end{equation}
which is usually referred to as Tsirelson bound. 

In order to reach this bound, we consider the maximally entangled state, that is we set $ \mathcal{C}=1 $ in \cref{eq:Psi-Schmidt}, leading to the singlet state 
\begin{equation}\label{eq:singlet-state}
	\ket{\Psi^{-}} = \frac{1}{\sqrt{2}}\left(\ket{0}_{A}\ket{1}_{B}-\ket{1}_{A}\ket{0}_{B}\right),
\end{equation} 
and further choose the measurement vectors such that $ \vec{q} \perp \vec{r} $ and $ \vec{s}\equiv -(\vec{q}+\vec{r})/\sqrt{2} $ as well as $ \vec{t}\equiv(\vec{q}-\vec{r})/\sqrt{2} $. Inserting these relations into \cref{eq:S-Kqrst}, and using that for $ \mathcal{C}=1 $, the correlation matrix $ K = -\mathds{1} $ is proportional to unity, we find 
\begin{equation}\label{key}
	\mathcal{S} = \frac{\abssq{\vec{q}}}{\sqrt{2}} + \frac{\abssq{\vec{r}}}{\sqrt{2}} = 2\sqrt{2},
\end{equation}
that is the Tsirelson bound, where we used that $ \vec{q} $ and $ \vec{r} $ are unit vectors. Note that by no means the conditions to maximally violate the CHSH inequality determines the measurement direction.

For a non-maximally entangled state, that is for $ \mathcal{C}<1 $, the Tsirelson bound can no longer be obtained. Furthermore, the measurement vectors that maximize $ \abs{\mathcal{S}} $ in general are no longer the same as the ones that correspond to the  Tsirelson bound. In the following we will discuss the optimal measurement strategy to obtain the maximum of $ \abs{\mathcal{S}} $, for a non-maximally entangled state, that is a state of the form of \cref{eq:Psi-Schmidt} with $ \mathcal{C} <1 $.

\section{Geometrical interpretation}\label{sec:geo}
\begin{figure}
	\centering
	\includegraphics{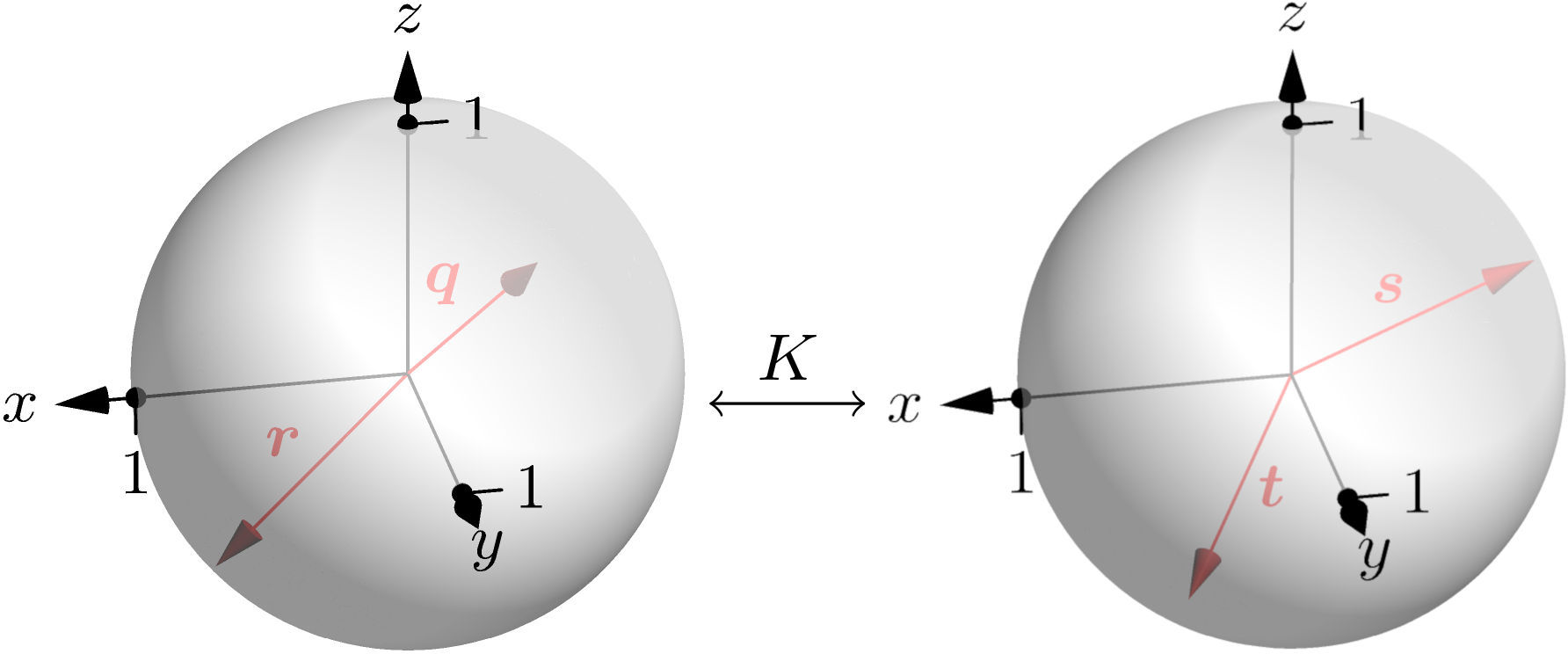}
	\caption{Geometric interpretation of the expectation value $ \mathcal{S} $ of the CHSH inequality in terms of \cref{eq:S-Kqrst}. The two pairs of vectors $ \vec{q} $ and $ \vec{r} $ as well as $ \vec{s} $ and $ \vec{t} $ are unit vectors which end on the Bloch spheres of the subsystems $ A $ and $ B $. The correlation matrix $ K $ establishes the connection between the vectors on the two Bloch spheres.}
	\label{fig:Blochspheres}
\end{figure}

The expectation value $ \mathcal{S} $ given by \cref{eq:S-Kqrst} has a geometrical interpretation shown in \cref{fig:Blochspheres}. The vectors $ \vec{q} $ and $ \vec{r} $ end on the Bloch sphere of the subsystem $ A $, while the vectors $ \vec{s} $ and $ \vec{t} $ terminate on the Bloch sphere of the subsystem $ B $. The expectation value $ \mathcal{S} $ contains all four possible pairwise combinations of one vector from subsystem $ A $ and one vector from the subsystem $ B $. The correlation matrix $ K $ defines these pairwise combinations, that is it defines how a vector from one subsystem is multiplied by a vector of the other subsystem. In a sense, it connects the two Bloch spheres with each other. 

\begin{figure}
	\centering
	\includegraphics{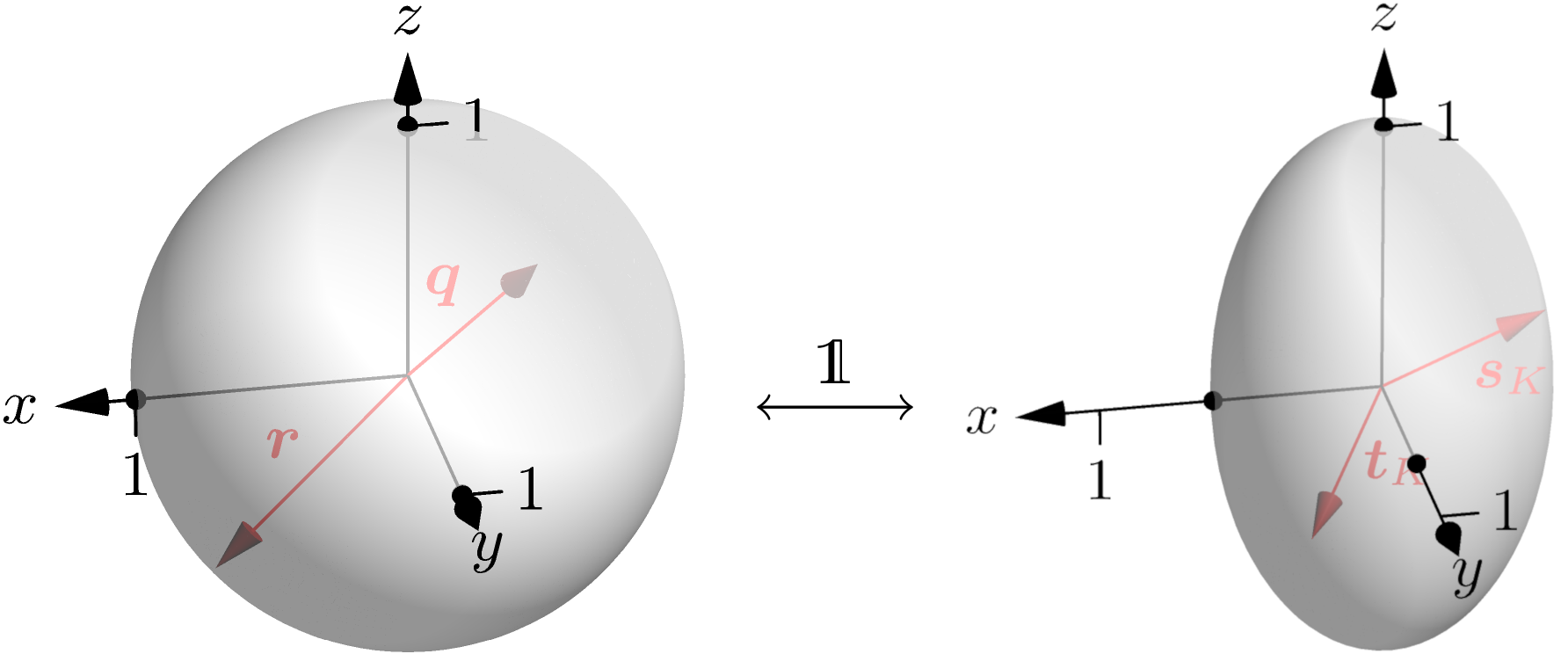}
	\caption{Elimination of the correlation matrix $ K $ in favour of the new vectors $ \vsk $ and $ \vtk $,  and geometrical interpretation of $ \mathcal{S} $ according to \cref{eq:S-primes}. Now $ \vec{q} $ and $ \vec{r} $ are still unit vectors, that is they still terminate on the Bloch sphere of the subsystem $ A $. However, due to the matrix $ K $ the new vectors $ \vec{s}_{K} $ and $ \vec{t}_{K} $ are no longer unit vectors, but vectors pointing to the surface of a prolate spheroid with the semi-minor axis given by the concurrence $ \mathcal{C} $. The Bloch sphere of subsystem $ A $ and the prolate spheroid of subsystem $ B $ are now connected by the identity matrix. }
	\label{fig:sphereandspheroid}
\end{figure}

Now we multiply the correlation matrix $ K $ by $ \vec{s} $ and $ \vec{t} $ of the subsystem $ B $, leading to new vectors 
\begin{equation}\label{eq:vsk}
	\vsk \equiv K \vec{s} 
\end{equation} 
and 
\begin{equation}\label{eq:vtk}
	\vtk \equiv K \vec{t}, 
\end{equation} 
which are no longer unit vectors. They do not end on the unit sphere anymore, but rather on a prolate spheroid with semi-axes of length $ \mathcal{C} $ and $ 1 $. This interpretation is depicted in \cref{fig:sphereandspheroid}.

When we insert the new vectors into \cref{eq:S-Kqrst} for $ \mathcal{S} $, and regroup the four terms, the expectation value
\begin{equation}\label{eq:S-primes}
	\mathcal{S} = \vec{q}\cdot\left(\vsk- \vtk\right) + \vec{r}\cdot\left(  \vsk + \vtk \right)
\end{equation}
is now only a combination of the scalar products between the vectors $ \vec{q} $ and $ \vec{r} $ from subsystem $ A $, and $ \vsk+\vtk $ and $ \vsk-\vtk $ of subsystem $ B $. 

We can simplify \cref{eq:S-primes} as
\begin{equation}\label{eq:S-abs1}
	\mathcal{S} = \abs{\vec{q}} \abs{\vsk- \vtk} \cos\alpha + \abs{\vec{r}} \abs{\vsk + \vtk} \cos\beta
\end{equation}
where $ \alpha $ and $ \beta $ are the angles between $ \vec{q} $ and $  \vsk - \vtk $ and between $ \vec{r} $ and $ \vsk + \vtk $, respectively. 
Since both, $ \vec{q} $ and $ \vec{r} $ are unit vectors, we find
\begin{equation}\label{eq:S-abs2}
	\mathcal{S} = \abs{\vsk- \vtk} \cos\alpha + \abs{\vsk + \vtk} \cos\beta.
\end{equation}

The vectors $ \vec{q} $ and $ \vec{r} $ can be chosen independently of each other and of the vectors $ \vsk $ and $ \vtk $. Therefore, the angles $ \alpha $ and $ \beta $ are also independent of each other and we can easily maximize over the vectors $ \vec{q} $ and $ \vec{r} $ by choosing them such that either $ \alpha = \beta = 0 $ or $ \alpha = \beta = \pi $. 

Therefore, we maximize $ \mathcal{S} $ by choosing $ \vec{q} $ and $ \vec{r} $ with $ \vsk $ and $ \vtk $
\begin{equation}\label{eq:vecq-ito-vskandvtk}
	\vec{q} \equiv \pm\frac{\vsk - \vtk}{\abs{\vsk -\vtk}}
\end{equation}
and 
\begin{equation}\label{eq:vecr-ito-vskandvtk}
	\vec{r} \equiv \pm\frac{\vsk + \vtk}{\abs{\vsk + \vtk}},
\end{equation}
with which we find from \cref{eq:S-abs2} the expression
\begin{equation}\label{eq:def-Sm}
	\mathcal{S}_{m}(\vsk,\vtk) \equiv \max_{\vec{q},\vec{r}}\abs{ \mathcal{S}} = \abs{\vsk- \vtk} + \abs{\vsk + \vtk}, 
\end{equation}
which only depends on vectors associated with the measurements performed on the subsystem $ B $ and the correlation matrix $ K $. The optimization problem is thereby reduced to optimizing the two vectors $ \vsk $ and $ \vtk $ on a spheroid.

\begin{figure*}
	\centering
	\includegraphics{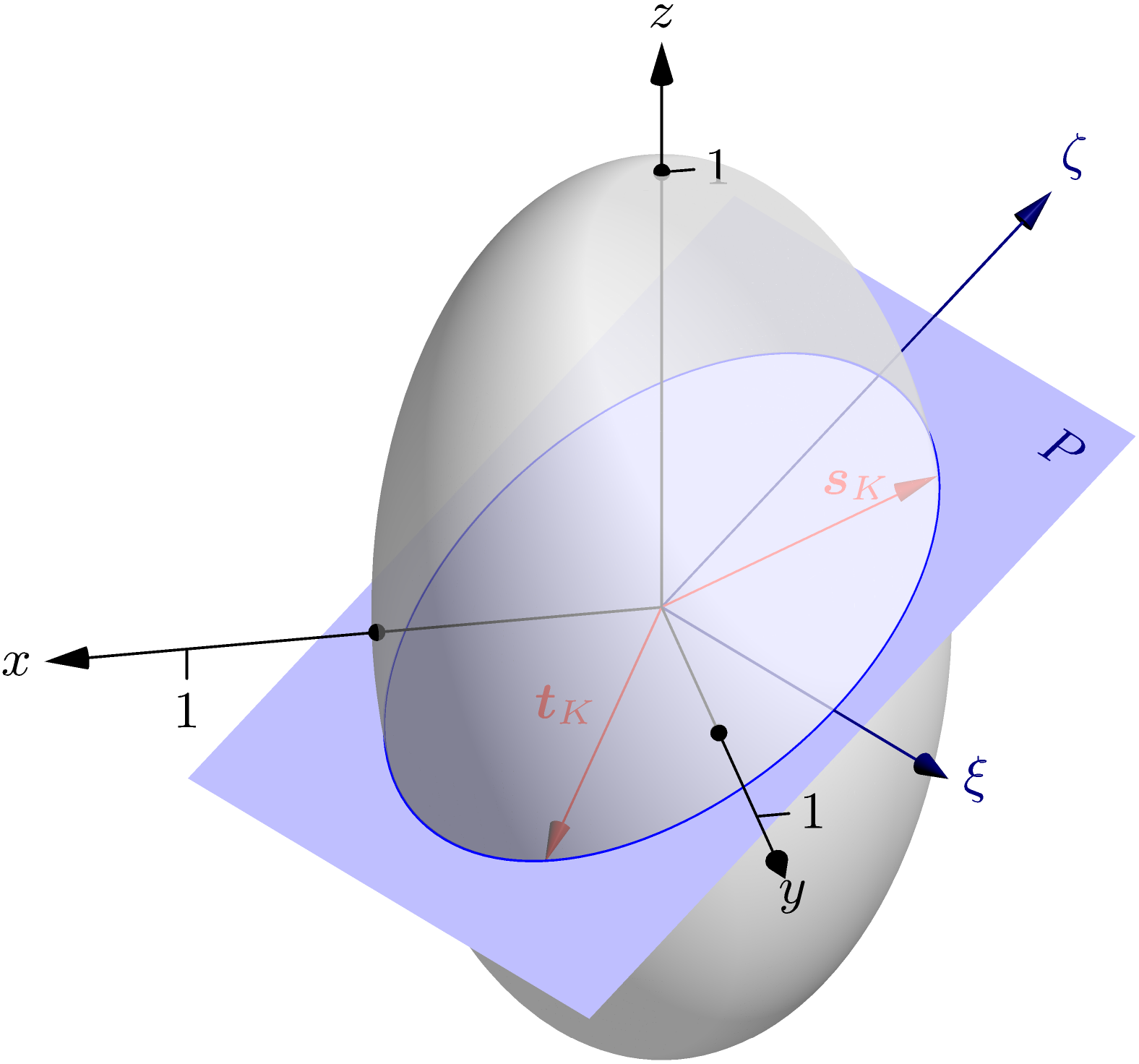}
	\includegraphics{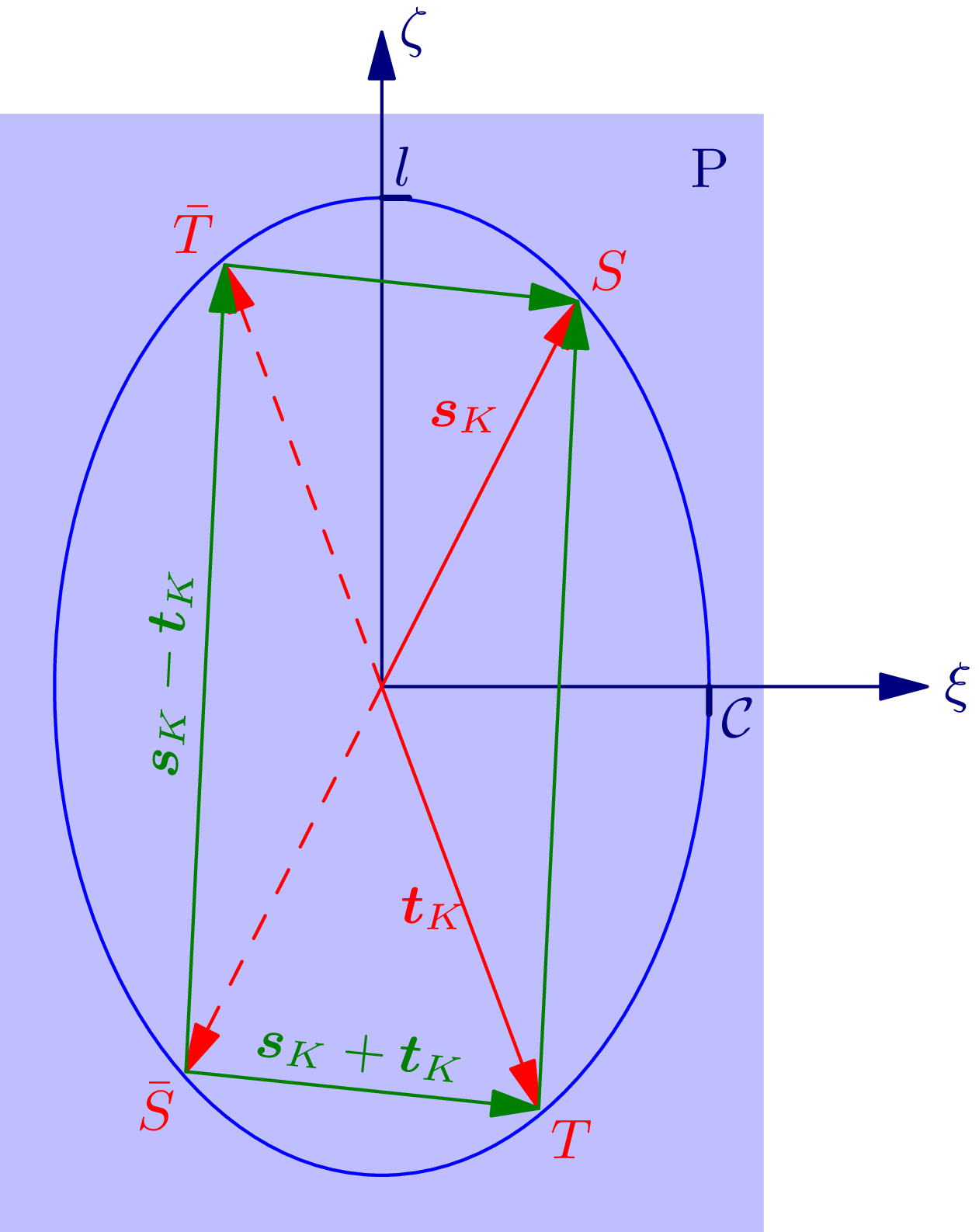}
	\caption{
		Geometrical interpretation of $ \mathcal{S} $ as the circumference of a parallelogram enclosed by an ellipse. We only consider the subsystem $ B $ of Fig. 2 represented on the left by the prolate spheroid. The vectors $ \vsk $ and $ \vtk $ span a plane $ P $ which cuts the spheroid in an ellipse with a semi-minor axis of length $ \mathcal{C} $ and a semi-major axis of length $ l $, with $ \mathcal{C}\leq l\leq 1 $, depending on the relative orientation of the plane to the ellipsoid. The vectors $ \vsk $ and $ \vtk $, which terminate on this ellipse, together with the vectors $ \vsk+\vtk $ and $ \vsk-\vtk $, which are also in $ P $, form a parallelogram whose corners $ S,T,\bar{S},\bar{T} $ lie on the ellipse. The task of maximizing the expectation value $ \mathcal{S}_{m}(\vsk,\vtk) $ with respect to $ \vsk $ and $ \vtk $ is therefore equivalent to maximizing the perimeter of the parallelogram.}
	\label{fig:EllipsoidandEllipse}
\end{figure*}

In general $ \vsk $ and $ \vtk $ span a two-dimensional plane $ P $, as shown in \cref{fig:EllipsoidandEllipse}. This plane contains both the center of the spheroid as well as the linear combinations $ \vsk + \vtk$ and $ \vsk - \vtk$. The intersection between the plane and the spheroid therefore is an ellipse. Since the intersection contains the origin of the spheroid, the resulting ellipse has a semi-minor axis of length $ \mathcal{C} $ and a semi-major axis of length $ l $, with $ \mathcal{C} \leq l \leq 1 $. 

Indeed, the exact value of the semi-major axis $ l $ depends on the relative orientation of the plane to the spheroid. If for example $ P $ is the $ x $-$ y $-plane, the intersection defines a circle with radius $ \mathcal{C} $, and we have $ l=\mathcal{C} $. On the other hand, if $ P $ contains the $ z $-axis, we have $ l =1 $. In all other cases, $ l $ is in between these two extreme cases. 

We now focus on the ellipse in the two-dimensional plane $P$ as depicted on the right-hand side of \cref{fig:EllipsoidandEllipse}. The vectors $ \vsk$ and $\vtk $ touch the ellipse in the points $ S $ and $ T $ respectively. The same is true for $ -\vsk $ and $ -\vtk $, and we call these points $ \bar{S} $ and $ \bar{T} $. Moreover, $ T $ and $ S $, as well as $ \bar{S} $ and $ \bar{T} $ are the connected by $ \vsk - \vtk $. Similarly, $ S $ and $ \bar{T} $, as well as $ \bar{S} $ and $ T $ are connected by $ \vsk +\vtk$. Two sides of the parallelogram formed by the four points $ S,T,\bar{S} $ and $ \bar{T} $ have the length $ \abs{\vsk-\vtk} $ and two have $ \abs{\vsk+\vtk} $. The perimeter of the resulting parallelogram is thus $ 2\mathcal{S}_{m} $. Maximizing the expectation value $ \mathcal{S}_{m} $ is therefore equivalent to finding the parallelogram enclosed by the ellipse with the maximal perimeter.

We note that the perimeter of the parallelogram increases when we increase the semi-axes of the ellipse. For a given state with a fixed concurrence $ \mathcal{C} $, the expectation value $ \mathcal{S}_{m} $ is maximal, when the semi-major axis is as long as possible, that is for $ l=1 $. Throughout the remainder of our article we set $ l=1 $. As discussed before, this choice implies that the plane $ P $ contains the $ z $-axis.

\section{Optimization of the measurement strategy}\label{sec:optimization}
We now determine measurement strategies that maximize the expectation value $ \mathcal{S}_{m} $. For this purpose, we first determine the angle between the vectors $ \vsk $ and $ \vtk $ that maximizes $ \mathcal{S}_{m} $. We then address the orientations of the vectors $ \vec{q},\vec{r},\vec{s}$ and $\vec{t} $ relative to each other. 

\subsection{A continuum of optimal angles} 
We start by expressing the three-dimensional unit vectors
\begin{equation}\label{eq:vecs-final1}
	\vec{s} =
	\begin{pmatrix}
		\cos\theta\cos\varphi_{s}\\
		\sin\theta\cos\varphi_{s}\\
		\sin\varphi_{s}
	\end{pmatrix} 
\end{equation}
and 
\begin{equation}\label{eq:vect-final1}
	\vec{t} =
	\begin{pmatrix}
		\cos\theta\cos\varphi_{t} \\
		\sin\theta\cos\varphi_{t} \\ \sin\varphi_{t}
	\end{pmatrix} ,
\end{equation}
in spherical coordinates with the polar angles $ \varphi_{s} $ and $ \varphi_{t} $, as well as the azimuth $ \theta $, which is the same for both vectors, since they span the plane $ P $ which contains the $ z $-axis. Note, that we have chosen the angles such that $ 0\leq \varphi_{s}, \varphi_{t} < 2\pi $ and $ 0\leq \theta < \pi $.  

When we insert these representations into the definitions, \cref{eq:vsk,eq:vtk}, of $ \vsk $ and $ \vtk $ as well as apply the explicit form, \cref{eq:K-diag}, of the correlation matrix $ K $ we find
\begin{equation}\label{eq:vecs-final2}
	\vec{s}_{K} = - 
	\begin{pmatrix}
		\mathcal{C}\cos\theta\cos\varphi_{s}\\
		\mathcal{C}\sin\theta\cos\varphi_{s}\\
		\sin\varphi_{s}
	\end{pmatrix} 
\end{equation}
and 
\begin{equation}\label{eq:vect-final2}
	\vec{t}_{K} = -
	\begin{pmatrix}
		\mathcal{C}\cos\theta\cos\varphi_{t} \\
		\mathcal{C}\sin\theta\cos\varphi_{t} \\ \sin\varphi_{t}
	\end{pmatrix}.
\end{equation}

Throughout the remainder of this section we choose a coordinate system in the two-dimensional plane $P$ defined by the variables $ \xi $ and $ \zeta $, such that the $ \xi $-axis points along the semi-minor axis, which lies in the $ x $-$ y $-plane, and the $ \zeta $-axis along the semi-major axis of the ellipse, that is the $\zeta$-axis coincides with the $z$-axis of the three-dimensional coordinate system.

In this two-dimensional system, we find the vector representations
\begin{equation}\label{eq:sprime-2d}
	\vsk = -
	\begin{pmatrix}
		\mathcal{C} \cos\varphi_{s}\\ \sin\varphi_{s}
	\end{pmatrix} 
\end{equation} and 
\begin{equation}\label{eq:tprime-2d}
	\vtk = -
	\begin{pmatrix}
		\mathcal{C}\cos\varphi_{t}\\ \sin\varphi_{t},
	\end{pmatrix}
\end{equation} 
which only depend on the polar angles $ \varphi_{s} $ and $ \varphi_{t} $ of the vectors $ \vec{s} $ and $ \vec{t} $ and the concurrence $ \mathcal{C} $. 

Inserting \cref{eq:sprime-2d,eq:tprime-2d} into \cref{eq:def-Sm}, we find the expression
\begin{align}\label{eq:Sm-angles-parameters}
	\mathcal{S}_{m} = &\sqrt{\mathcal{C}^{2}(\cos\varphi_{s}-\cos\varphi_{t})^{2}+(\sin\varphi_{s}-\sin\varphi_{t})^{2}} \\ \nonumber +
	&\sqrt{\mathcal{C}^{2}(\cos\varphi_{s}+\cos\varphi_{t})^{2}+(\sin\varphi_{s}+\sin\varphi_{t})^{2}}
\end{align}
for the expectation value $ \mathcal{S}_{m} $, as a function of $ \varphi_{s} $ and $ \varphi_{t} $, which is already maximized over $ \vec{q} $ and $ \vec{r} $ by the choice made in \cref{eq:vecq-ito-vskandvtk,eq:vecr-ito-vskandvtk}. 

\begin{figure}
	\centering
	\includegraphics{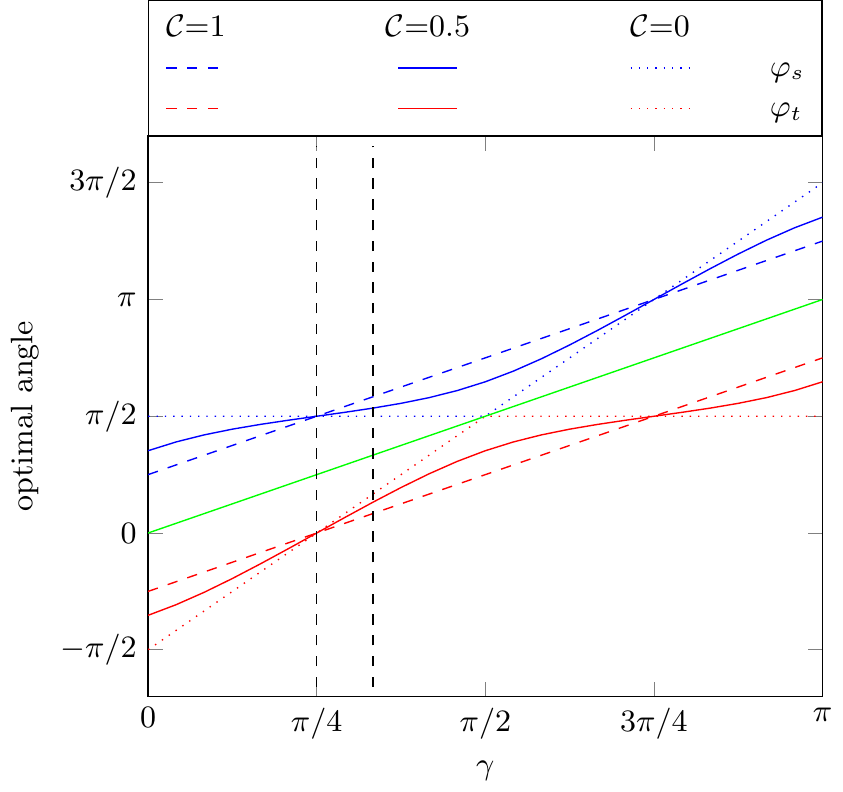}
	\caption{Dependence of the optimal polar angles $ \varphi_{s} $ and $ \varphi_{t} $ on the  angle $ \gamma $ maximizing the expectation value $ \mathcal{S} $ for the three different concurrences $ \mathcal{C}=1 $ (dashed), $ \mathcal{C}=0.5 $ (solid), and $ \mathcal{C}=0 $ (dotted). For a maximally entangled state (dashed parallel lines), the difference between the two angles is always $ \varphi_{s} -\varphi_{t} = \pi/2 $, that is the measurements are orthogonal on each other. For a non-maximally entangled state (solid lines), the difference between the optimal angles depends on $ \gamma $. For a separable state (dotted lines), the difference between optimal angles are either $ \varphi_{s}=\pi/2 $ or $ \varphi_{t}=\pi/2 $. The average $ (\varphi_{s}+\varphi_{t})/2 $ of the two angles is $ \gamma $ as given by \cref{eq:maximal-angles-pair,eq:maximal-angles-pair2} and shown by the green solid line. The vertical dashed lines mark the special values of $ \gamma $ used in \cref{fig:optimal-measurement-vectors}. }
	\label{fig:optimalphist}
\end{figure} 

In Appendix \ref{app:derivation}, we analytically maximize $ \mathcal{S}_{m} $ and find the remarkable property that there exist not a single, but infinitely many possible combinations of $ \varphi_{s} $ and $ \varphi_{t} $, which can be parameterized by the average angle 
\begin{equation}\label{eq:gamma}
	\gamma \equiv (\varphi_{s} + \varphi_{t})/2 
\end{equation}
as 
\begin{equation}\label{eq:maximal-angles-pair}
	\varphi_{s}(\gamma) = \gamma + \delta
\end{equation}
and 
\begin{equation}\label{eq:maximal-angles-pair2}
	\varphi_{t}(\gamma) =  \gamma - \delta
\end{equation}
with the difference angle
\begin{equation}\label{eq:delta-explicit}
	\delta =  \arccot\sqrt{\frac{\mathcal{C}^{2}\cos^{2}\gamma +\sin^{2}\gamma}{\mathcal{C}^{2}\sin^{2}\gamma +\cos^{2}\gamma}}
\end{equation}
This result implies, that there exist infinitely many parallelograms which are inscribed by an ellipse with identical perimeters. This interesting property of ellipses and parallelograms has been shown before in References \cite{Richard2004,Connes2007}, where also a geometric proof has been given.

In \cref{fig:optimalphist} we show the dependence of the optimal angles $ \varphi_{s} $ and $ \varphi_{t} $ on $ \gamma $ for different values of $ \mathcal{C} $, based on \cref{eq:maximal-angles-pair,eq:maximal-angles-pair2}.

For $ \mathcal{C}=1 $, we get from \cref{eq:maximal-angles-pair,eq:maximal-angles-pair2}
\begin{equation}\label{key}
	\varphi_{s}(\gamma)= \gamma + \frac{\pi}{4}
\end{equation}
and 
\begin{equation}\label{key}
	\varphi_{t}(\gamma) = \gamma - \frac{\pi}{4}.
\end{equation}
Hence, the only requirement to maximize the expectation value $ \mathcal{S} $ is that the two vectors $ \vec{s} $ and $ \vec{t} $ are perpendicular, which is shown in \cref{fig:optimalphist}. This result is in agreement with the literature \cite{Clauser1969}.

For $ \mathcal{C} < 1 $, $ \vec{s}_{K} $ and $ \vec{t}_{K} $ are no longer perpendicular for all average angles $ \gamma $. Instead, for $ \gamma < \pi/4 $ or $ \gamma > 3\pi/4 $, the angle between $ \vec{s}_{K} $ and $ \vtk $ is obtuse, increasing with decreasing concurrence, while for $ \pi/4 < \gamma <3\pi/4 $ the angle between $ \vsk $ and $ \vtk $ is acute and decreases with decreasing concurrence.

In the limit of $ \mathcal{C} = 0 $, for $ \gamma < \pi/2 $ the angle $ \varphi_{s} = \pi/2 $ is constant. From the definition, \cref{eq:gamma}, of the average angle $ \gamma $ it directly follows that $ \varphi_{t} = 2\gamma -\pi/2 $ depends linearly on $ \gamma $. Since we have chosen $ \varphi_{s} \geq \varphi_{t} $, for $ \gamma > \pi/2 $ the angle $ \varphi_{t} = \pi/2 $ is constant and $ \varphi_{s}=2\gamma-\pi/2 $ depends linearly on $ \gamma $.   

\subsection{A parallelogram enclosed in an ellipse}
We now use the optimal angles $ \varphi_{s} $ and $ \varphi_{t} $ to determine the value of the maximum of the correlation measure $ \mathcal{S} $, and interpret the results in terms of our geometrical picture. 

When we substitute the expression for $ \varphi_{s} $ and $ \varphi_{t} $, \cref{eq:maximal-angles-pair,eq:maximal-angles-pair2}, into \cref{eq:Sm-angles-parameters} we find that the expectation value $ \mathcal{S}_{m} $ assumes its maximal value 
\begin{equation}\label{eq:maxS-ito-C}
	\max \abs{\mathcal{S}}=\max_{\vsk,\vtk} \mathcal{S}_{m}(\vsk,\vtk) = 2\sqrt{1+\mathcal{C}^{2}}.
\end{equation} 

\begin{figure*}
	\centering
	\includegraphics{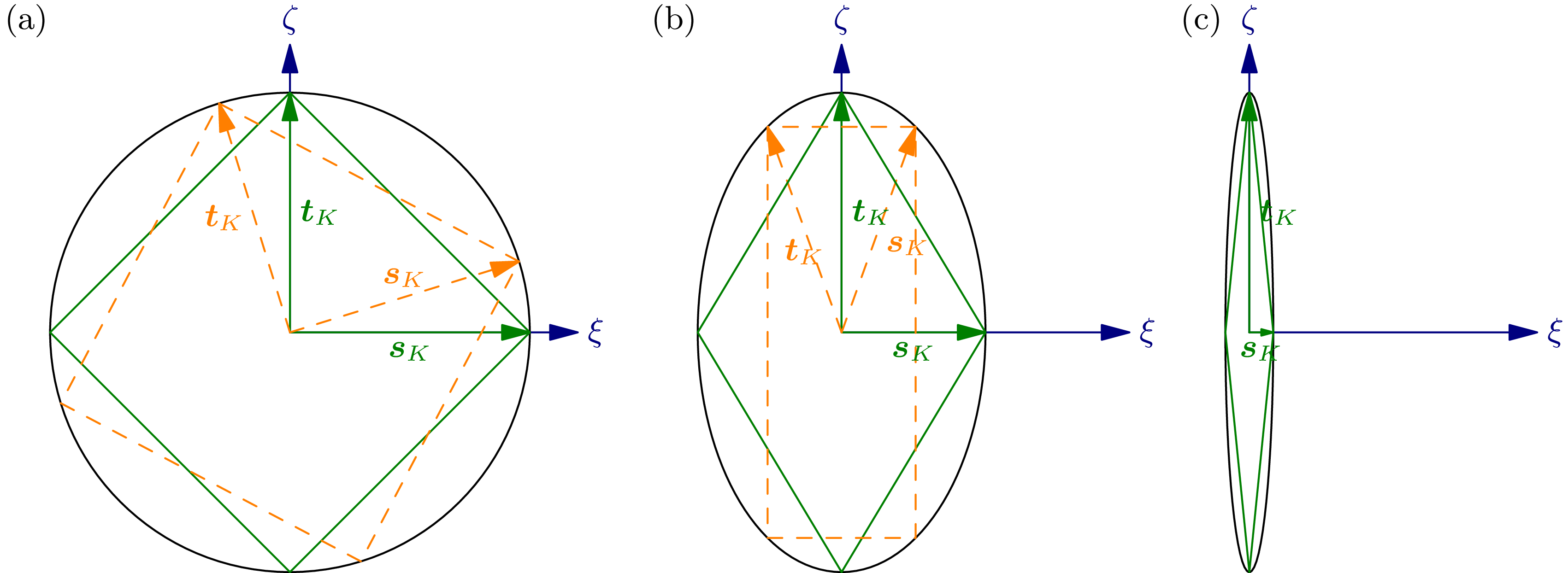}
	\caption{Expectation value $ \mathcal{S} $ as the perimeter of a parallelogram for different concurrences $ \mathcal{C} $ based on Equation (15). (a) For a maximally entangled state, that is for $ \mathcal{C}=1 $, the ellipse reduces to a unit circle and the parallelogram with the maximal perimeter is a square with edge length $ \sqrt{2} $. Due to the rotational symmetry of the circle, the green square can be rotated inside the circle without changing the perimeter as exemplified by the dashed orange square. (b) For $ 0 < \mathcal{C} < 1 $, the shape of the optimal parallelogram depends on the direction of one of the vectors $ \vsk $ or $ \vtk $. Despite the different shapes the perimeters are identical. (c) As $ \mathcal{C} $ approaches zero, the ellipse becomes more and more elongated along the $ \zeta $-axis. The optimal measurement is achieved for either $ \vsk $ or $ \vtk $ being almost parallel to this axis and the other one orthogonal. In the limit of $ \mathcal{C} =0$ the parallelogram reduces to a line along the $ \zeta $-axis. }
	\label{fig:maximalSforC=1andC=0}
\end{figure*}

In the case of $ \mathcal{C}=1 $, that is for a maximally entangled state, \cref{eq:maxS-ito-C} reduces to the well-known result \cite{Cirelson1980} \begin{equation}\label{key}
	\max \abs{\mathcal{S}} = 2\sqrt{2}
\end{equation}
and for $ \mathcal{C}=0 $, we get the mere classical bound \cite{Clauser1969} \begin{equation}\label{key}
	\max \abs{\mathcal{S}} = 2 .
\end{equation} 
Note that however small the entanglement of the pure state is, with the suitable measurement one can in principle violate the CHSH inequality.

We now discuss our results in the geometrical picture of a parallelogram inscribed by an ellipse. We show this behaviour for three different concurrences $ \mathcal{C} $ in \cref{fig:maximalSforC=1andC=0}.

For $ \mathcal{C}=1 $ the ellipse reduces to a unit circle, as depicted in \cref{fig:maximalSforC=1andC=0}(a). In this case, the parallelogram assumes its maximal perimeter when it describes a square whose sides have length $ \sqrt{2} $. Since the circle is invariant under rotations about its center, the square can also be rotated about this point without changing its perimeter. Therefore, all squares inscribed by the circle are possible solutions.

For $ 0 <\mathcal{C}<1 $ the geometry is more complicated, since there is no rotational symmetry anymore. The parallelograms associated with the optimal measurement strategy are no longer squares, and not even necessarily rectangles. In \cref{fig:maximalSforC=1andC=0}(b), we present two parallelograms for maximizing the perimeter. 

In the first example, shown in green, the vectors $ \vsk $ and $ \vtk $ point along the coordinate axes $ \xi $ and $ \zeta $, that is the parallelogram connects the semi-minor and the semi-major axes of the ellipse, and becomes a rhombus, whose perimeter is $ 4\sqrt{1+\mathcal{C}^{2}} $. 

The second realization depicted in \cref{fig:maximalSforC=1andC=0}(b) in orange is the optimal rectangle. This shape is optimal, when the tangents on the ellipse at the endpoints of $ \vsk $ and $ \vtk $ are perpendicular to each other \cite{Connes2007}. This orthogonality condition is true for all possible solutions, as we show in Appendix \ref{app:orthogonality}, and thus allows a geometric construction of all possible optimal parallelograms. We note, that all other solutions in general have less symmetry, and can be found from \cref{eq:maximal-angles-pair,eq:maximal-angles-pair2}.

As $ \mathcal{C} $ approaches $ 0 $, leading to a separable state $ \ket{\Psi} $, the ellipse collapses to the $ \zeta $-axis, as suggested in \cref{fig:maximalSforC=1andC=0}(c). The parallelogram therefore is maximal if one of the two vectors $ \vsk $ and $ \vtk $ coincides with the semi-major axis, while the other vector is orthogonal to it. The parallelogram degenerates into a line between $ \zeta=-1 $ and $ \zeta=1 $ on the $ \zeta $-axis. It therefore has a perimeter of $ 2(1-(-1))=4 $, leading to $ \max\mathcal{S}_{m}=2 $.

\subsection{Optimal measurement vectors}
In the previous section, we have derived the optimal two-dimensional vectors $ \vsk $ and $ \vtk $.
We now construct the four optimal three-dimensional measurement vectors $ \vec{q},\vec{r},\vec{s} $ and $ \vec{t} $. 
%

For this purpose we obtain the vectors $ \vec{s} $ and $ \vec{t} $ by inserting the optimal angles $ \varphi_{s} $ and $ \varphi_{t} $ given by \cref{eq:maximal-angles-pair,eq:maximal-angles-pair2} into the spherical representations, \cref{eq:vecs-final1,eq:vect-final1}, of $ \vec{s} $ and $ \vec{t} $. They represent a family of measurement directions on the Bloch sphere of subsystem $B$ parameterized by the average angle $ \gamma $ and the free parameter $ \theta $. 

Next we recall that the optimal measurement vectors $ \vec{q} $ and $ \vec{r} $ on the subsystem $ A $ are given by \cref{eq:vecq-ito-vskandvtk,eq:vecr-ito-vskandvtk}. When we insert the explicit expressions, \cref{eq:vecs-final2,eq:vect-final2}, of $ \vsk $ and $ \vtk $, together with the optimal angles, \cref{eq:maximal-angles-pair,eq:maximal-angles-pair2}, into the definitions, \cref{eq:vecq-ito-vskandvtk,eq:vecr-ito-vskandvtk}, of $ \vec{q} $ and $ \vec{r} $ we find the representations
\begin{equation}\label{eq:vecq-res1}
	\vec{q}(\gamma)=\mp\frac{1}{\sqrt{\mathcal{C}^{2}\sin^{2}\gamma + \cos^{2}\gamma}}
	\begin{pmatrix}
		-\mathcal{C}\cos\theta\sin\gamma\\
		-\mathcal{C}\sin\theta\sin\gamma\\
		\cos\gamma
	\end{pmatrix}
\end{equation} 
and 
\begin{equation}\label{eq:vecr-res1}
	\vec{r}(\gamma)=\mp\frac{1}{\sqrt{\mathcal{C}^{2}\cos^{2}\gamma + \sin^{2}\gamma}}
	\begin{pmatrix}
		\mathcal{C}\cos\theta\cos\gamma\\
		\mathcal{C}\sin\theta\cos\gamma\\ 
		\sin\gamma
	\end{pmatrix} .
\end{equation}
Note that when for instance $ \vec{q}\parallel \vec{s} $ then the actual measurement directions are not necessarily parallel as in the Schmidt representation of a given state $ \ket{\Psi} $ the local three-dimensional coordinate systems in general are rotated mutually.

Since $ \vec{q} $ and $ \vec{r} $ are by definition unit vectors, we can rewrite them in spherical coordinates as
\begin{equation}\label{eq:vecq-res2}
	\vec{q}(\gamma) = \mp\begin{pmatrix}
		\cos\theta \cos\varphi_{q}(\gamma)\\
		\sin\theta \cos\varphi_{q}(\gamma)\\
		\sin\varphi_{q}(\gamma)
	\end{pmatrix}
,
	\vec{r}(\gamma) = \mp\begin{pmatrix}
		\cos\theta \cos\varphi_{r}(\gamma)\\
		\sin\theta \cos\varphi_{r}(\gamma)\\
		\sin\varphi_{r}(\gamma)
	\end{pmatrix}
\end{equation}
with the $ \gamma $-dependent polar angles
\begin{equation}\label{eq:varphi_q}
	\varphi_{q}(\gamma) \equiv \arctan\left(-\frac{1}{\mathcal{C}}\cot\gamma\right) + {\pi}
\end{equation}
and 
\begin{equation}\label{eq:varphi_r}
	\varphi_{r}(\gamma) \equiv \arctan\left(\frac{1}{\mathcal{C}}\tan\gamma\right) + \Theta\left(\gamma - \frac{\pi}{2}\right)\pi.
\end{equation}
Here, $ \Theta(x) $ denotes the Heaviside step-function and we assume $ -\pi/2 \leq \arctan x \leq \pi/2 $. 

The constant $ \pi $ in \cref{eq:varphi_q,eq:varphi_r} ensures that \cref{eq:vecq-res1,eq:vecq-res2}, as well as \cref{eq:vecr-res1} describe the same vectors, respectively.

\begin{figure}
	\centering
	\includegraphics{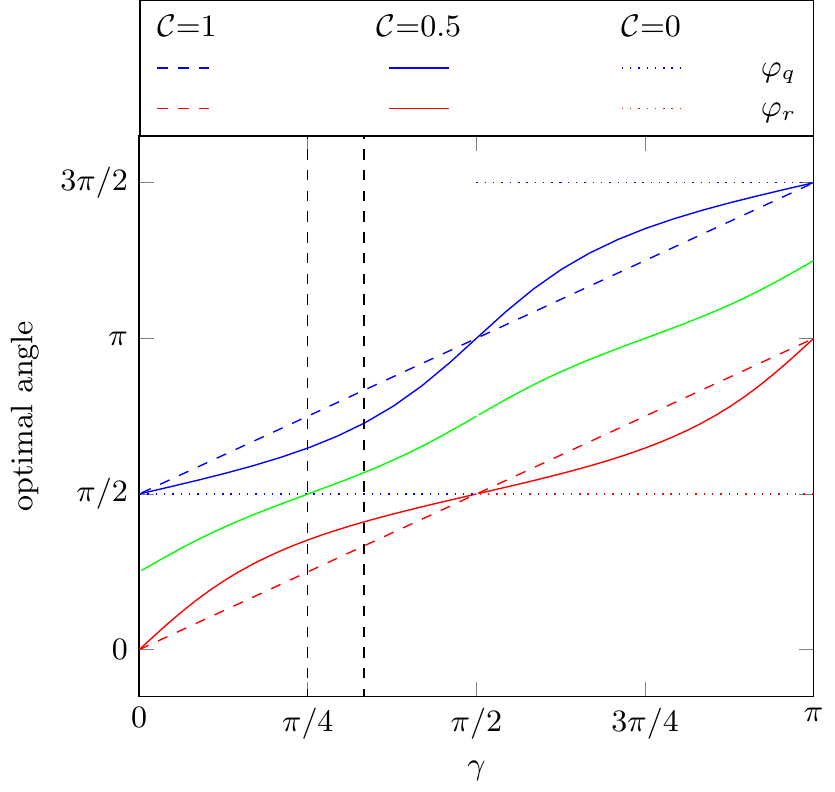}
	\caption{Dependence of the optimal polar angles $ \varphi_{q} $ and $ \varphi_{r} $ on the angle $ \gamma $ maximizing the expectation value $ \mathcal{S} $ for three different values of the concurrence $ \mathcal{C} $. For a maximally entangled state, that is $ \mathcal{C}=1 $, (dashed lines) the difference between the two angles is always $ \pi/2 $. For $ \mathcal{C}<1 $ the difference between the angles $ \varphi_{q} $ and $ \varphi_{r} $ decreases in the regime of $ \gamma < \pi/2 $ and increases for $ \gamma >\pi/2 $. In contrast to \cref{fig:optimalphist}, the average $ (\varphi_{q}+\varphi_{r})/2 $ of the two angles does not linearly depend on $ \gamma $ for $ \mathcal{C}=0.5 $ (green solid line). The vertical dashed lines mark the special values of $ \gamma $ used in \cref{fig:optimal-measurement-vectors}.}
	\label{fig:optimalangles-q-r}
\end{figure}
In \cref{fig:optimalangles-q-r} we present the dependence of $ \varphi_{q} $ and $ \varphi_{r} $ on the parameter $ \gamma $. For $ 0\leq \gamma <\pi/2 $, the difference between the two angles $ \varphi_{q} $ and $ \varphi_{r} $ \textit{decreases} with decreasing concurrence $ \mathcal{C} $. In the regime of $ \pi/2 < \gamma < \pi $, the angle difference \textit{increases} with a decreasing entanglement. 

When we compare this dependence of the angles $ \varphi_{q} $ and $ \varphi_{r} $ on $ \gamma $, with the dependence of $ \varphi_{s} $ and $ \varphi_{t} $ on the same parameter, as shown in \cref{fig:optimalphist}, we see that their dependence on the concurrence is shifted by an angle $ \pi/4 $. 
Furthermore, while $ \gamma $ is the angle bisection of the angle between $ \vec{s} $ and $ \vec{t} $, this is not true for the angle between $ \vec{q} $ and $ \vec{r} $. Indeed, the green line of \cref{fig:optimalangles-q-r} depends on $ \gamma $.  


\subsection{Special examples}
We conclude our discussions by considering three different examples that maximize the expectation value $ \mathcal{S} $. For simplicity, we only consider the positive solutions of the vectors $ \vec{q} $ and $ \vec{r} $ in \cref{eq:vecq-res2}. 

\begin{figure}
	\centering
	\includegraphics{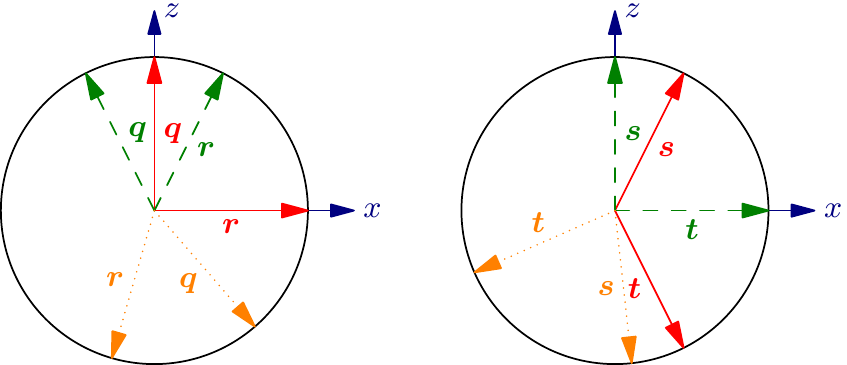}
	\caption{Optimal measurement vectors $ \vec{q} $, $ \vec{r} $, $ \vec{s} $ and $ \vec{t} $ for three different values of $ \gamma $, but for a fixed concurrence $ \mathcal{C}=0.5 $. For  $ \gamma = 0 $ (red solid vectors) $ \vec{q} $ and $ \vec{r} $ are orthogonal, as indicated by \cref{fig:optimalangles-q-r}, while the angle between the vectors $ \vec{s} $ and $ \vec{t} $ is maximal, as depicted in \cref{fig:optimalphist}. In the case of $ \gamma = \pi/4 $ (green dashed vectors),  $ \vec{s} $ and $ \vec{t} $ is orthogonal, while the angle between $ \vec{q} $ and $ \vec{r} $ is minimal. For $ \gamma = 4\pi/3 $ (orange dotted) the vectors in neither subsystem are orthogonal.}
	\label{fig:optimal-measurement-vectors}
\end{figure}

In \cref{fig:optimal-measurement-vectors} we depict the optimal vectors $ \vec{q} $, $ \vec{r} $, $ \vec{s} $ and $ \vec{t} $ for these three different choices of the parameter $ \gamma $. Here, we always choose a fixed azimuth $ \theta=0 $, hence the vectors lie in the $ x $-$ z $-plane. In the following, we therefore provide only the $ x $- and $ z $-components of the vectors, and suppress the $ y $-component, which always vanishes. 

The most elementary example is $ \gamma = 0 $, where the vectors read
\begin{equation}\label{eq:final-vectors-special-case}
	\vec{q}= \begin{pmatrix}
		0\\1
	\end{pmatrix}, 
	\vec{r}= \begin{pmatrix}
		1\\0
	\end{pmatrix},
\end{equation}
and 
\begin{equation}\label{eq:final-vectors-special-case1}
	\vec{s}= \frac{1}{\sqrt{1+\mathcal{C}^{2}}}\begin{pmatrix}
		\mathcal{C}\\1
	\end{pmatrix},  \vec{t}= \frac{1}{\sqrt{1+\mathcal{C}^{2}}}\begin{pmatrix}
		\mathcal{C}\\-1
	\end{pmatrix}.
\end{equation}
This result shows again that the relative angle between the measurements depend on the entanglement on the state. For $ \mathcal{C}=1 $, it reduces to the well known settings \cite{Clauser1969}, but for $ \mathcal{C} <1 $ the angle $ 2\delta $ between the vectors $ \vec{s} $ and $ \vec{t} $ is implicitly given by 
\begin{equation}\label{key}
	\cos 2\delta = -\frac{1-\mathcal{C}^{2}}{1+\mathcal{C}^{2}},
\end{equation}
which is negative since $ 0 < \mathcal{C} <1 $. Therefore, this angle is always greater than $ \pi/2 $, and increases with decreasing concurrence $ \mathcal{C} $.

A similar situation is found for $ \gamma = \pi/4 $. Here, the vectors read
\begin{equation}\label{eq:final-vectors-special-case2}
	\vec{q}= \frac{1}{\sqrt{1+\mathcal{C}^{2}}}\begin{pmatrix}
		-\mathcal{C}\\1
	\end{pmatrix},\qquad \vec{r}= \frac{1}{\sqrt{1+\mathcal{C}^{2}}}\begin{pmatrix}
		\mathcal{C}\\1
	\end{pmatrix},
\end{equation}
and
\begin{equation}\label{eq:final-vectors-special-case21}
	\vec{s}= \begin{pmatrix}
		0\\1
	\end{pmatrix}, \qquad
	\vec{t}= \begin{pmatrix}
		1\\0
	\end{pmatrix},
\end{equation}
that is now $ \vec{s} $ and $ \vec{t} $ of the subsystem $ B $ are orthogonal, while the angle $ \eta $ between the vectors $ \vec{q} $ and $ \vec{r} $ is given by 
\begin{equation}\label{key}
	\cos\eta = \frac{1-\mathcal{C}^{2}}{1+\mathcal{C}^{2}} .
\end{equation} 
In contrast to the previous example $ \gamma = 0 $, this angle is smaller than $ \pi/2 $ for $ \mathcal{C}<1 $, and decreases with decreasing concurrence $ \mathcal{C} $.

The last example in \cref{fig:optimal-measurement-vectors} is $ \gamma = 4\pi/3 $, for which we obtain
\begin{equation}\label{eq:final-vectors-special-case3}
	\vec{q}= \frac{1}{\sqrt{1+3\mathcal{C}^{2}}}\begin{pmatrix}
		\sqrt{3}\mathcal{C}\\-1
	\end{pmatrix}, \quad \vec{r}= \frac{1}{\sqrt{3+\mathcal{C}^{2}}}\begin{pmatrix}
		-\mathcal{C}\\-\sqrt{3}
	\end{pmatrix}, 
\end{equation}
for the subsystem $ A $ and 
\begin{equation}\label{eq:final-vectors-special-case32}
	\vec{s}= \begin{pmatrix}
		\cos\left(\frac{4\pi}{3} + \phi_{0}\right)\\ \sin\left(\frac{4\pi}{3} + \phi_{0}\right)
	\end{pmatrix}, \quad
	\vec{t}= \begin{pmatrix}
		\cos\left(\frac{4\pi}{3} - \phi_{0}\right)\\ \sin\left(\frac{4\pi}{3} - \phi_{0}\right)
	\end{pmatrix}, 
\end{equation}
for the subsystem $ B $ with the concurrence-dependent phase
\begin{equation}\label{key}
	\phi_{0} = \arctan\sqrt{\frac{1+3\mathcal{C}^{2}}{3+\mathcal{C}^{2}}},
\end{equation}
which in general has to be calculated numerically. 

In this case, neither the vectors $ \vec{q} $ and $ \vec{r} $ of the subsystem $ A $, nor the vectors $ \vec{s} $ and $ \vec{t} $ of $ B $ are orthogonal for $ \mathcal{C}<1 $. This example illustrates the existence of solutions that are not easy to guess.


\section{Mixed states}\label{sec:mixed}
So far, we only considered the optimization of the correlation measure \eqref{eq:S-avg-Operators} for pure two-qubit states. We turn to the problem of mixed two-qubit states. 

In order to do so, we generalize the correlation matrix $ K $, \cref{eq:K-operators}, defined for pure states by the expression
\begin{equation}\label{eq:quantum-corr:mixed-def-K}
	K \equiv \tr(\hat{\rho}_{AB}\cdot\hat{\vec{\sigma}}_{A}\otimes\hat{\vec{\sigma}}_{B})
\end{equation}
to two-qubit mixed states $ \hat{\rho}_{AB} $. 

For pure states, we demonstrated in Section \ref{sec:geo}, that the maximization of the expectation value $ \mathcal{S} $ has a geometrical interpretation as the maximization of the circumference of a parallelogram inscribed by an ellipse. The deeper origin of the ellipse is the transformation of the Bloch sphere of one subsystem by the correlation matrix $ K $ into a prolate spheroid. This transformation directly follows from the diagonal form of the correlation matrix, \cref{eq:K-diag}, of any pure state.

The diagonal form of the correlation matrix for pure states was a consequence of writing the state in the Schmidt basis. Unfortunately, we cannot generalize this property directly to the mixed state case. 
The question now is, if such a geometrical interpretation is still possible for the mixed two-qubit states, since it is not clear if the mixed state again transforms the Bloch sphere into a spheroid and our geometrical argument still holds for mixed states. 

In this section, we first show that any mixed state in fact transforms the Bloch sphere of subsystem $ B $ into an ellipsoid, which in general is less symmetric than the prolate spheroid of the pure state case. We then exploit this result to show that a similar geometrical interpretation exists as for the pure state case, and derive the corresponding optimal angles and maximal expectation value $ \mathcal{S} $. We conclude this section by considering two examples.

\subsection{Transformed Bloch sphere: Still an ellipsoid}
We show in the following, that the correlation matrix $ K $ transforms the Bloch sphere of subsystem $ B $ into an ellipsoid with three in general different semi-axes $ a $, $ b $ and $ c $, with $ a\geq b \geq c $. In addition, the orientation of the ellipsoid with respect to our coordinate system and the lengths $ a $, $ b $ and $ c $ depends on the state $ \hat{\rho}_{AB} $. 

We start by considering an arbitrary vector $ \vec{v} $ which points from the origin to the surface of the Bloch sphere of subsystem $ B $. As a consequence the vector $ \vec{v} $ is a unit vector. The correlation matrix $ K $ then defines a new vector
\begin{equation}\label{eq:appB2:vk-def}
	\vec{v}_{K} \equiv K \vec{v}.
\end{equation} 

For any correlation matrix $ K $, there exists a singular value decomposition \cite{Nielsen2001}
\begin{equation}\label{key}
	K=U\Sigma V^{T} 
\end{equation}
with two orthogonal matrices $ U $ and $ V $ and the diagonal matrix \begin{equation}\label{key}
	\Sigma=\diag(a,b,c) ,
\end{equation} 
such that $ a\geq b \geq c \geq 0 $. 

We note that, since $ \abs{\det K} = abc $ all elements of the diagonal matrix $ \Sigma $ are strictly positive if the inverse of$ K $ exists, while at least one element vanishes if $ K $ is not invertible.

For simplicity we choose $ a,b,c>0 $, that is all singular values to be positive. The case of at least one vanishing singular values can be discussed analogously. 

When we insert the singular value decomposition into the definition, \cref{eq:appB2:vk-def}, of $ \vec{v}_{K} $ we obtain
\begin{equation}\label{key}
	\vec{v}_{K} = U\Sigma V^{T}  \vec{v}
\end{equation} 
for the new vector.

We can interpret the action of the three matrices $ U\Sigma V^{T}  $ as following: First, we rotate the coordinate system by the orthogonal transformation $ V $. Then the matrix $ \Sigma $ deforms the Bloch sphere along the axes of the new coordinate system by scaling it with the respective singular value. Finally, the orthogonal matrix $ U $ rotates the new shape. 

When we introduce the abbreviations $ \vec{v}_{K}^{\prime} \equiv U^{T}\vec{v}_{K} $ and $ \vec{v}^{\prime} \equiv V^{T} \vec{v} $, that is directly perform the rotations of the two orthogonal matrices on the two vectors $ \vec{v}_{K} $ and $ \vec{v} $, respectively, we get
\begin{equation}\label{key}
	\vec{v}_{K}^{\prime} = \Sigma  \vec{v}^{\prime}.
\end{equation}

As a next step, we introduce the inverse
\begin{equation}\label{eq:inverse-Sigma}
	\Sigma^{-1} = \diag(a^{-1},b^{-1},c^{-1})
\end{equation}
of the diagonal matrix $ \Sigma $, which leads us to 
\begin{equation}\label{key}
	\Sigma^{-1}\vec{v}_{K}^{\prime} =  \vec{v}^{\prime}.
\end{equation}
When we take the absolute square of the vectors on both sides, we find 
\begin{equation}\label{key}
	\vec{v}_{K}^{\prime T}(\Sigma^{-1})^{T}\Sigma^{-1}\vec{v}_{K}^{\prime} =  1.
\end{equation}
Here, we made use of the norm $ \abssq{\vec{v}^{\prime}} = \abssq{\vec{v}}=1 $, which is a direct consequence of $ \vec{v} $ being a unit vector by definition, as it points from the center of the Bloch sphere to its surface, and the fact that orthogonal matrices do not change the norm, such that $ \vec{v}^{\prime} $ is also a unit vector. 

When we further use the relation 
\begin{equation}\label{key}
	(\Sigma^{-1})^{T}\Sigma^{-1} = \diag(a^{-2},b^{-2},c^{-2}) ,
\end{equation} 
we finally arrive at
\begin{equation}\label{key}
	\left(\frac{{v}_{K,1}^{\prime}}{a}\right)^{2}+ \left(\frac{{v}_{K,1}^{\prime}}{b}\right)^{2}+ \left(\frac{{v}_{K,3}^{\prime}}{c}\right)^{2} = 1,
\end{equation}
which is the equation of an ellipsoid with semi axes $ a $, $ b $ and $ c $. 


We finally relate this result back to the vectors $ \vec{v} = V\vec{v}^{\prime} $ and $ \vec{v}_{K} = U\vec{v}_{K}^{\prime} $. Since the two matrices $ U $ and $ V $ are both orthogonal, the vectors $ \vec{v} $ and $ \vec{v}_{K} $ are only rotations of the vectors $ \vec{v}^{\prime} $ and $ \vec{v}_{K}^{\prime} $, respectively. Therefore, any vector $ \vec{v} $ pointing from the center of the Bloch sphere to its surface is mapped onto an ellipsoid with semi-axes of length $ a $, $ b $ and $ c $. The orientation of this ellipsoid with respect to the original coordinate system is defined by the matrices $ U $ and $ V $. Hence, the family of states with the same singular values $ a $, $ b $ and $ c $ transforms the Bloch sphere into ellipsoids of identical shape but different orientation. 


In our derivation, we only considered all singular values to be non-vanishing. In full analogy to the above derivation, we can also consider the case of vanishing singular values. The only difference is that for vanishing singular values the inverse $ \Sigma^{-1} $, \cref{eq:inverse-Sigma}, has to be replaced by a pseudo-inverse. Consequently, the dimensions of the problem are reduced. As a result, for one singular value, instead of an ellipsoid, the Bloch sphere is transformed into an area enclosed by an ellipse, while for two vanishing singular values the Bloch sphere is transformed into a line. In the case of all three singluar values being vanishing it is trivial to see that the Bloch sphere is mapped onto the origin of the coordinate system, that is onto a point.

\subsection{Geometry and maximization}
In the preceding section we have shown that the correlation matrix transforms the Bloch sphere into an ellipsoid. The vectors $ \vsk $ and $ \vtk $ point from the origin to the surface of this ellipsoid.

The intersection of the plane $ P $, spanned by the vectors $ \vsk $ and $ \vtk $, and the new ellipsoid is still an ellipse. The optimal plane $ P $, that is the plane which leads to the ellipse with the largest semi axes, creates an ellipse with semi-major axis $ a $ and semi-minor axis $ b $. We note that for a general ellipsoid there exists only one distinct plane $ P $, which cuts the ellipsoid into this optimal ellipse. In contrast, for a prolate spheroid there exist infinitely many such planes, due to the rotational symmetry of the spheroid. Therefore, going from a pure state to a non-pure mixed state breaks this symmetry.  

The parallelogram inscribed by the ellipse is a result of the measurement vectors alone, and does not change when changing the state of the system. Therefore, we now have a similar geometrical interpretation of maximizing the circumference of a parallelogram, inscribed by an ellipse. The only difference relevant for the present discussion between the pure and the mixed state cases therefore lies in the different length of the semi axes of the ellipse. 

The optimization strategy is thus the same as for the pure state case, discussed in Section \ref{sec:optimization}. The only difference is that instead of semi-axes of length one and $ \mathcal{C} $, we find $ a $ and $ b $. As a result, the optimal angles are still given by \cref{eq:maximal-angles-pair,eq:maximal-angles-pair2}, however, the difference angle $ \delta $, \cref{eq:delta-explicit}, has to change.
We recall from the definition of $ \delta $, \cref{eq:delta-explicit}, that for an ellipse with semi-major axis of length $ l $ and semi-minor axis of length $ \mathcal{C} $, the angle is given by  
\begin{equation}\label{eq:delta-explicit-mixed-lC}
	\delta =  \arccot\sqrt{\frac{\mathcal{C}^{2}\cos^{2}\gamma +l^{2}\sin^{2}\gamma}{\mathcal{C}^{2}\sin^{2}\gamma +l^{2}\cos^{2}\gamma}},
\end{equation}   
where we have originally set $ l=1 $. The extension to an arbitrary value of $ l $, can be easily made from the argument that the optimal angles $ \varphi_{s} $ and $ \varphi_{t} $, and thus the difference angle $ \delta $, depend only on the ratio $\mathcal{C}/l $ of the two semi-axes, but not on their absolute values. Indeed, stretching both axes by a constant factor will not change the optimal angles, but only the perimeter of the associated parallelogram. 

Thus, when we make the substitution $ \mathcal{C} \to b$ and $ \l \to a $ in the difference angle $ \delta $, \cref{eq:delta-explicit-mixed-lC}, we find the new difference angle 
\begin{equation}\label{eq:delta-explicit-mixed}
	\delta =  \arccot\sqrt{\frac{b^{2}\cos^{2}\gamma +a^{2}\sin^{2}\gamma}{b^{2}\sin^{2}\gamma +a^{2}\cos^{2}\gamma}},
\end{equation}
for the family of optimal angles $ \varphi_{s} = \gamma +\delta $ and $ \varphi_{t} = \gamma -\delta $, parameterized by the average angle $ \gamma $. 

All that is left to do, is to find the maximal expectation value $ \mathcal{S} $. In order to do so, we recall from our discussion in Section \ref{sec:optimization}, that one of the optimal parallelograms is the rhombus whose corners are determined by the intersection of the ellipse and its semi-axes (confer the green parallelogram in \cref{fig:maximalSforC=1andC=0}(b)). By using the Pythagorean theorem, it is straightforward to see that each edge of this rhombus has length $ \sqrt{a^{2}+b^{2}} $. 
Therefore the maximal correlation measure is 
\begin{equation}\label{eq:maxS-mixed-state}
	\max \abs{\mathcal{S}} = 2\sqrt{a^{2}+b^{2}},
\end{equation} 
since it is half of the perimeter of the parallelogram. 

We note that from \cref{eq:maxS-mixed-state} it immediately follows that any mixed state can violate the CHSH inequality, \cref{eq:CHSH-inequality}, if and only if $ a^{2}+b^{2}>1 $. Our result is in full agreement with the criterion found in Ref. \cite{Horodecki1995} on the violation of the CHSH inequality for any mixed two-qubit state $ \hat{\rho} $.

\subsection{Mixture of entangled states}
In the following, we consider an example of mixed two-qubit states. We apply the general concepts developed in the previous paragraphs in order to deduce the optimal correlation measures $ \mathcal{S} $ and the optimal measurements for this states.

We consider the mixture
\begin{equation}\label{eq:rho-example-two}
	\hat{\rho} = p \ket{\Psi}\bra{\Psi} + (1-p)\ket{\Phi}\bra{\Phi}
\end{equation}
of the pure state  
\begin{equation}\label{eq:Psi-example}
	\ket{\Psi} = \sqrt{\frac{1+\sqrt{1-\mathcal{C}^{2}}}{2}}\ket{0}_{A}\ket{1}_{B} - \sqrt{\frac{1-\sqrt{1-\mathcal{C}^{2}}}{2}}\ket{1}_{A}\ket{0}_{B}
\end{equation}
with concurrence $ \mathcal{C} $, that occurs with probability $ p $, and 
\begin{equation}\label{eq:Phi-example}
	\ket{\Phi} = \sqrt{\frac{1+\sqrt{1-\mathcal{D}^{2}}}{2}}\ket{0}_{A}\ket{0}_{B} + \sqrt{\frac{1-\sqrt{1-\mathcal{D}^{2}}}{2}}\ket{1}_{A}\ket{1}_{B}.
\end{equation}
with concurrence $ \mathcal{D} $, which has probability $ 1-p $. 

By inserting the state $ \rho $, \cref{eq:rho-example-two}, into the definition of the correlation matrix, \cref{eq:quantum-corr:mixed-def-K}, together with the definitions of the individual basis states, \cref{eq:Psi-Schmidt,eq:Phi-example}, we find the correlation matrix 
\begin{equation}\label{eq:K-example-two-states-mixture}
	K = 
	\begin{pmatrix}
		-p\mathcal{C} + (1-p)\mathcal{D} & 0 & 0 \\
		0 & -p\mathcal{C} - (1-p)\mathcal{D} & 0 \\
		0 & 0 & 1-2p
	\end{pmatrix}
\end{equation}
of the mixed state $ \hat{\rho} $, which is already diagonal. Thus, the transformation due to this matrix describes an ellipsoid with in general three different semi-axes, that are oriented along the $ x $-, $ y $- and $ z $-coordinate. 

The lengths of these three semi-axes are the absolute values of the three diagonal entries of the correlation matrix, \cref{eq:K-example-two-states-mixture}. We note that the three axes in general have different lengths. Furthermore, the semi-axis along the $ z $-coordinate is no longer of unit length, as for the pure state case. In fact, for $ p=1/2 $ the axis even has vanishing length. This result originates from the different sign that the individual correlation matrices, \cref{eq:K-operators}, of the pure basis states $ \ket{\Psi} $ and $ \ket{\Phi} $ have in the $ zz $-component.  

\begin{figure}
	\centering
	\includegraphics{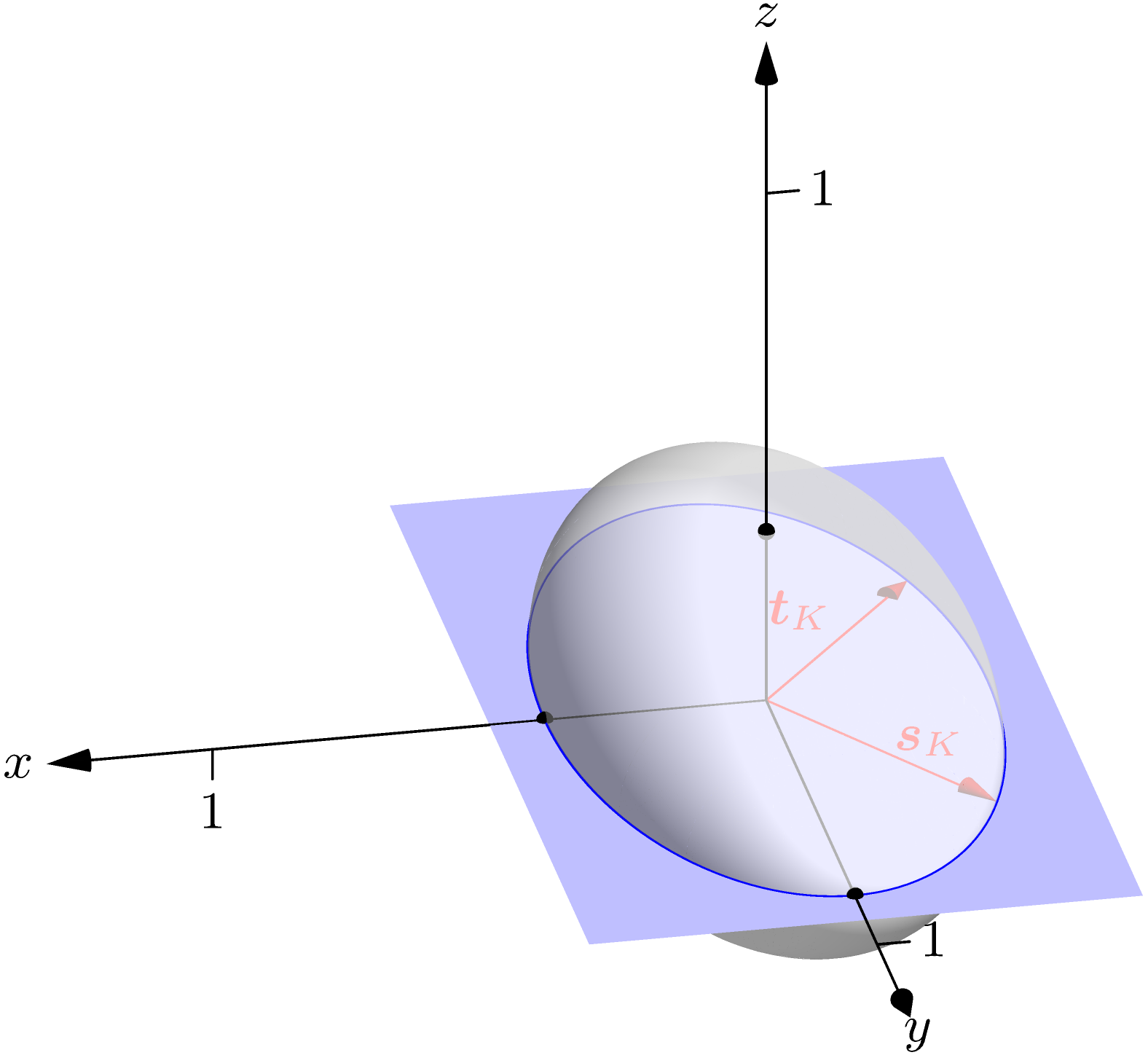}
	\caption{Geometrical interpretation of the correlation matrix $ K $ for the mixed state given by \cref{eq:rho-example-two}, for $ p=1/3 $, $ \mathcal{C} = 0.6 $ and $ \mathcal{D} =0.9 $. The correlation matrix is an ellipsoid with three different semi-axes along the $ x $, $ y $ and $ z $-coordinate, instead of a prolate spheroid of a pure state (cf. \cref{fig:EllipsoidandEllipse}). The intersection of the spheroid with the plane $ P $ spanned by the vectors $ \vsk $ and $ \vtk $ is still an ellipse. The maximal correlation measure can be obtained when $ P $ cuts the ellipsoid in the ellipse with the largest semi-axes. In this case, the optimal plane $ P $ is the $ x $-$ y $-plane. }
	\label{fig:mixed-ellipsoid}
\end{figure}
In \cref{fig:mixed-ellipsoid}, we show the ellipsoid created by \cref{eq:K-example-two-states-mixture} for the choice $ p=1/3 $, $ \mathcal{C} = 0.6 $ and $ \mathcal{D} =0.9 $. 

In order to find the maximal value of the correlation measure $ \mathcal{S} $, we need to find the plane $ P $ spanned by the vectors $ \vsk $ and $ \vtk $, that maximizes the ellipse when cutting the ellipsoid. 

Since $ \abs{p\mathcal{C}+(1-p)\mathcal{D}}\geq\abs{p\mathcal{C}-(1-p)\mathcal{D}} $, the $ y $-axis of the ellipsoid is always longer than its $ x $-axis. Furthermore, if $ \abs{p\mathcal{C}+(1-p)\mathcal{D}}\geq 1-2p $, the $ x $-axis is longer than the $ z $-axis, and the optimal plane is the $ x $-$ y $-plane. This behaviour is in contrast to the pure state case, where the optimal plane $ P $ always contained the $ z $-axis. Otherwise, if the $ z $-axis is longer than the $ x $-axis, the optimal plane is the $ y $-$ z $-plane. 

In the example depicted in \cref{fig:mixed-ellipsoid}, we find the optimal ellipse when the vectors $ \vsk $ and $ \vtk $ lie in the $ x $-$ y $-plane. This ellipse has a semi-major axis $ a=0.8 $ and $ b=0.4 $. From \cref{eq:maxS-mixed-state}, we immediately find the maximal expectation value $ \mathcal{S} = 2\sqrt{4/5} <2 $. Therefore, this mixed state can never violate the CHSH inequality, even though it is composed of two maximally entangled states. 

We note, that the optimal measurement vectors $ \vec{q} $, $ \vec{r} $, $ \vec{s} $ and $ \vec{v} $ are found by the same techniques as for the pure state case, discussed in Section \ref{sec:optimization}. In contrast to the pure state case, where we had the angle $ \theta $ as a remaining degree of freedom, due to the rotational symmetry of the spheroid, for an ellipsoid the optimal plane is in general completely determined. 

We conclude our discussion of the mixed state case with the remark, that in our examples the correlation matrix $ K $ is already diagonal. For a general mixed state, an important step is to first find the diagonal representation of the matrix $ KK^{T} $, in order to obtain the principal axes of the ellipsoid. A non-diagonal matrix $ K $ leads to an ellipse whose semi-axes are not given by the coordinate axes anymore. Once these axes are known, the maximal expectation value $ \mathcal{S} $ can easily be determined by \cref{eq:maxS-mixed-state}, and the optimal measurement vectors are obtained by using the results of Section \ref{sec:optimization}. We note that for a pure state this diagonalization is automatically done by using the Schmidt decomposition of the state. 

\subsection{Entanglement and violation of the CHSH inequality for mixed states}
As a second example of entangled two-qubit mixed states, we consider the famous Werner state \cite{Werner1989}
\begin{equation}\label{eq:Werner_state}
	\hat{\rho}_{W} \equiv p \ket{\Psi^{(-)}}\bra{\Psi^{(-)}} + \frac{1-p}{4}\idone,
\end{equation}
which is a mixture of the maximally entangled singlet state $ \ket{\Psi^{(-)}} $, as defined in \cref{eq:singlet-state}, with probability $ p $, and the completely mixed state $ \idone/4 $ in four dimensions with probability $ 1-p $. 
When we insert \cref{eq:Werner_state} into the definition of the correlation matrix, \cref{eq:quantum-corr:mixed-def-K}, we directly obtain
\begin{equation}\label{key}
	K = \diag(-p,-p,-p) = -p\mathds{1},
\end{equation}
which transforms the Bloch sphere of subsystem $ B $ into a smaller sphere with radius $ p $. This transformation is not surprising, since the correlation matrix for the singlet state $ \ket{\Psi^{-}} $ does not change the shape of the Bloch sphere, while the completely mixed state transforms the Bloch sphere into a single point at the origin of the sphere. Thus, the convex combination of these two actions still preserves the rotational symmetry of the sphere, but with a decreased radius. 

When we apply our geometrical considerations, we see that the plane $ P $ cuts the sphere into a circle with radius $ p $. The optimal measurement strategies are the same as for the maximally-entangled pure state, and from \cref{eq:maxS-mixed-state}, we directly obtain   
\begin{equation}\label{key}
	\mathcal{S} \leq 2\sqrt{2}p
\end{equation}
for the maximal value of the correlation function. Hence, the Werner state allows to violate the CHSH inequality for $ p > 1/\sqrt{2} $. 

From the PPT-criterion \cite{Peres1996,Horodecki1996} it is known, that the Werner state is non-separable for $ p>1/3 $. Hence, for $ 1/3<p<1/\sqrt{2} $ the state is entangled but cannot violate the CHSH-inequality. This result is in full agreement with the literature \cite{Horodecki1996}, and in contrast to any pure state, which violates CHSH for an appropriate measurement if and only if it is entangled.

\section{Conclusions and outlook} \label{sec:conclusionsandoutlook}
In this article we have demonstrated that the quantum mechanical expectation value $\mathcal{S}$ measuring the correlation in the CHSH inequality for a two-qubit entangled pure state can be interpreted geometrically as half of the perimeter of a parallelogram with its corners on an ellipse whose semi-minor axis is determined by the entanglement of the pure bipartite state and whose circumference is maximal. The larger the entanglement characterized by the concurrence $\mathcal{C}$, the larger is the semi-minor axis. 

Moreover, this interpretation also allowed us to analytically find the optimal measurement directions for any pure entangled two-qubit state. We demonstrated that there exist infinitely many such directions.

Finally, we extended our considerations to mixed two-qubit states. The main difference compared to the pure state case are the resulting correlation matrices $ K $, which in turn determine the curves enclosing the parallelogram. These curves remain ellipses, and the same geometrical interpretation as for pure states is possible. In contrast to the pure state case, the semi-major axes of the ellipse is no longer of unit length, and the semi-minor axes is not determined by the concurrence anymore.

An remaining open question is, whether our geometrical interpretation can also be applied to other types of Bell inequalities or similar equalities. In 1993 L. Hardy \cite{Hardy1993} introduced a test of non-classicality of two qubits that exploits an equality instead of an inequality. Interestingly, this equality is applicable only for all non-maximally entangled states, which indicates that a lack of symmetry in the underlying state is necessary. Therefore, it would be an interesting task to check, whether a similar geometrical interpretation exists for this equality and how it brings out this symmetry argument.     

\section*{Acknowledgments}
We are grateful to M. Freyberger, A. Friedrich, H. Losert, B. Pacolli and A. Wolf for many fruitful discussions.
J.S. thanks the Center for Integrated Quantum Science  and  Technology  (IQ\textsuperscript{ST})  for  a  fellowship within the framework of the Quantum Alliance sponsored by the Ministry of Science, Research and Arts, Baden-W{\"u}rttemberg. T.S. acknowledges support from the EU Quantum Flagship project QRANGE (grant no. 820405). W.P.S. is grateful to Texas A{\&}M University for a Faculty Fellowship at the Hagler Institute for Advanced Study at Texas A{\&}M University and to Texas A{\&}M AgriLife Research for the support of this work. The research of IQ\textsuperscript{ST} is financially supported by the Ministry of Science, Research and Arts, Baden-W{\"u}rttemberg.

\appendix
\section{Analytical derivation of the maximal violation}\label{app:derivation}
In this appendix we analytically find the angles $ \varphi_{s} $ and $ \varphi_{t} $ which maximize the expectation value
\begin{align}\label{eq:app:Sm-angles-parameters}
	\mathcal{S}_{m} = &\sqrt{\mathcal{C}^{2}(\cos\varphi_{s}-\cos\varphi_{t})^{2}+(\sin\varphi_{s}-\sin\varphi_{t})^{2}} \\ \nonumber +
	&\sqrt{\mathcal{C}^{2}(\cos\varphi_{s}+\cos\varphi_{t})^{2}+(\sin\varphi_{s}+\sin\varphi_{t})^{2}}.
\end{align}
For this purpose we introduce the average angle
\begin{equation}\label{eq:app:gamma}
	\gamma \equiv (\varphi_{s}+\varphi_{t})/2 
\end{equation}
and the difference angle 
\begin{equation}\label{eq:app:delta}
	\delta \equiv (\varphi_{s}-\varphi_{t})/2 ,
\end{equation} yielding the expression
\begin{align}\label{eq:Sm-gamma-delta}
	\mathcal{S}_{m} = 2\abs{\sin\delta}\sqrt{\mathcal{C}^{2}\sin^{2}\gamma + \cos^{2}\gamma} \nonumber \\+
	2\abs{\cos\delta} \sqrt{\mathcal{C}^{2}\cos^{2}\gamma+\sin^{2}\gamma}.
\end{align}
For the sake of simplicity we restrict ourselves to the interval $ 0\leq \delta \leq \pi/2 $. This constraint can always be fulfilled, since for $ \delta >\pi/2 $, we can simply exchange the two vectors $ \vsk $ and $ \vtk $.

We are now able to maximize this expression by first calculating the gradient, that is the derivatives
\begin{align}\label{eq:partialSgamma}
	\frac{\partial \mathcal{S}_{m}}{\partial \gamma} = & 2\sin\delta\frac{\mathcal{C}^{2}\cos\gamma\sin\gamma-\sin\gamma\cos\gamma}{\sqrt{\mathcal{C}^{2}\sin^{2}\gamma + \cos^{2}\gamma}} \nonumber \\  &-
	2\cos\delta\frac{\mathcal{C}^{2}\sin\gamma\cos\gamma-\sin\gamma\cos\gamma}{\sqrt{\mathcal{C}^{2}\cos^{2}\gamma+\sin^{2}\gamma}} = 0
\end{align}
and 
\begin{align}\label{eq:partialSdelta}
	\frac{\partial \mathcal{S}_{m}}{\partial \delta} = & 2\cos\delta\sqrt{\mathcal{C}^{2}\sin^{2}\gamma + \cos^{2}\gamma} \nonumber \\ &-	2\sin\delta \sqrt{\mathcal{C}^{2}\cos^{2}\gamma+\sin^{2}\gamma} = 0
\end{align}
of the expectation value and then equating both to zero. 

In this way we find from \cref{eq:partialSdelta} the relation
\begin{equation}\label{eq:app:4}
	\cos\delta = \sin\delta \frac{\sqrt{\mathcal{C}^{2}\cos^{2}\gamma+\sin^{2}\gamma}}{\sqrt{\mathcal{C}^{2}\sin^{2}\gamma + \cos^{2}\gamma}},
\end{equation}
which by inserting into \cref{eq:partialSgamma} leads us to
\begin{widetext}
\begin{equation}\label{eq:app:5}
	\sin\delta\frac{\mathcal{C}^{2}\cos\gamma\sin\gamma-\sin\gamma\cos\gamma}{\sqrt{\mathcal{C}^{2}\sin^{2}\gamma + \cos^{2}\gamma}} +
	\sin\delta \frac{\sqrt{\mathcal{C}^{2}\cos^{2}\gamma+\sin^{2}\gamma}}{\sqrt{\mathcal{C}^{2}\sin^{2}\gamma + \cos^{2}\gamma}}\cdot\frac{-\mathcal{C}^{2}\sin\gamma\cos\gamma+\sin\gamma\cos\gamma}{\sqrt{\mathcal{C}^{2}\cos^{2}\gamma+\sin^{2}\gamma}} =0.
\end{equation}
\end{widetext}
Hence, a vanishing derivative of $ \mathcal{S}_{m} $ with respect to $ \delta $ also implies a vanishing derivative of $ \mathcal{S}_{m} $ with respect to $ \gamma $. In particular, \cref{eq:app:5} 
is fulfilled for all $ \gamma $ and $ \delta $. 

Therefore, for every $ \gamma $ we can choose according to \cref{eq:app:4}
\begin{equation}\label{eq:delta-optimal}
	\delta = \arccot\sqrt{\frac{\mathcal{C}^{2}\cos^{2}\gamma + \sin^{2}\gamma}{\mathcal{C}^{2}\sin^{2}\gamma + \cos^{2}\gamma}},
\end{equation}
and the gradient of $ \mathcal{S}_{m} $ vanishes.

When we insert this expression for $ \delta $ into \cref{eq:Sm-gamma-delta}, the expectation value becomes
\begin{equation}\label{key}
	\max_{\vec{s},\vec{t}} \mathcal{S}_{m} = 2\sqrt{1+\mathcal{C}^{2}},
\end{equation}
independent of the parameter $ \gamma $. 

Hence, for every $ \gamma $, we can find a parameter $ \delta $, such that the expectation value assumes its global maximum.  We can finally invert the coordinate transformation, \cref{eq:app:gamma,eq:app:delta}, by $ \varphi_{s} = \gamma + \delta $ and $ \varphi_{t} = \gamma -\delta $, which together with \cref{eq:delta-optimal} defines the optimal measurements. 

\section{Orthogonality of the tangents on the ellipse}\label{app:orthogonality}
In this appendix we show that the optimal parallelograms in terms of maximizing the expectation value $ \mathcal{S} $ are geometrically determined by the constraint that the tangents on the ellipse in the  points $ S $ and $ T $ are orthogonal. In order to do so, we first note that the tangent vectors are determined by the derivatives 
\begin{equation}\label{key}
	\frac{\mathrm{d}\vsk}{\mathrm{d}\varphi_{s}} = \begin{pmatrix}
		\mathcal{C} \sin\varphi_{s}\\
		-\cos\varphi_{s}
	\end{pmatrix}
\end{equation}
and 
\begin{equation}\label{key}
	\frac{\mathrm{d}\vtk}{\mathrm{d}\varphi_{t}} = \begin{pmatrix}
		\mathcal{C} \sin\varphi_{t}\\
		-\cos\varphi_{t}
	\end{pmatrix}
\end{equation}
of the vectors $ \vsk $ and $ \vtk $, defined by \cref{eq:sprime-2d,eq:tprime-2d}, respectively, leading to the scalar product
\begin{equation}\label{eq:app2:scalar-prod}
	\frac{\mathrm{d}\vsk}{\mathrm{d}\varphi_{s}}\cdot \frac{\mathrm{d}\vtk}{\mathrm{d}\varphi_{t}} = \mathcal{C}^{2}\sin\varphi_{s}\sin\varphi_{t} + \cos\varphi_{s}\cos\varphi_{t}.
\end{equation}

When we insert the optimal angles $ \varphi_{s} = \gamma + \delta  $ and $ \varphi_{t} = \gamma - \delta $, defined by \cref{eq:maximal-angles-pair,eq:maximal-angles-pair2} into \cref{eq:app2:scalar-prod}, we find
\begin{equation}\label{eq:app2:scalar-prod2}
	\frac{\mathrm{d}\vsk}{\mathrm{d}\varphi_{s}}\cdot \frac{\mathrm{d}\vtk}{\mathrm{d}\varphi_{t}} = \mathcal{C}^{2}\sin(\gamma+\delta)\sin(\gamma-\delta) + \cos(\gamma+\delta)\cos(\gamma-\delta),
\end{equation}
which with help of the trigonometric relations \begin{equation}\label{key}
	\cos(x+y)\cos(x-y) = \cos(2x) + \cos(2y) 
\end{equation} 
and 
\begin{equation}\label{key}
	\sin(x+y)\sin(x-y) = \cos(2y) - \cos(2x) , 
\end{equation}  
simplifies to
\begin{equation}\label{eq:app2:scalar-prod3}
	\frac{\mathrm{d}\vsk}{\mathrm{d}\varphi_{s}}\cdot \frac{\mathrm{d}\vtk}{\mathrm{d}\varphi_{t}} = \frac{1+\mathcal{C}^{2}}{2}\cos(2\delta)+\frac{1-\mathcal{C}^{2}}{2}\cos(2\gamma).
\end{equation}
From the definition of $ \delta $, \cref{eq:delta-explicit}, and the relation \begin{equation}\label{key}
	\cos(2\arccot x)=\frac{x^{2}-1}{x^{2}+1} 
\end{equation} we find 
\begin{equation}\label{eq:app2-cos2delta}
	\cos(2\delta) = \frac{\mathcal{C}^{2}(\cos^{2}\gamma-\sin^{2}\gamma) +(\sin^{2}\gamma - \cos^{2}\gamma)}{\mathcal{C}^{2}(\cos^{2}\gamma+\sin^{2}\gamma) +(\sin^{2}\gamma + \cos^{2}\gamma)},
\end{equation}
which when we make use of the relations 
\begin{equation}\label{key}
	 \cos^{2}\gamma + \sin^{2}\gamma = 1 
\end{equation} 
and 
\begin{equation}\label{key}
	 \cos^{2}\gamma - \sin^{2}\gamma = \cos(2\gamma) 
\end{equation}
leads to
\begin{equation}\label{eq:app2-delta2d-final}
	\cos(2\delta) =  -\frac{1-\mathcal{C}^{2}}{1+\mathcal{C}^{2}}\cos(2\gamma).
\end{equation}

When we finally insert \cref{eq:app2-delta2d-final} into the scalar product \cref{eq:app2:scalar-prod3}, we find 
\begin{equation}\label{key}
	\frac{\mathrm{d}\vsk}{\mathrm{d}\varphi_{s}}\cdot \frac{\mathrm{d}\vtk}{\mathrm{d}\varphi_{t}}  = 0,
\end{equation}
proving the orthogonality of the two tangent vectors.

	\end{document}